\documentclass[twocolumn,floats,showpacs,superscriptaddress,pre]{revtex4}
\usepackage{amssymb}
\usepackage{graphicx}
\usepackage{dcolumn}
\usepackage{amsmath}
\usepackage{bm}
\usepackage{epsfig}

\def\sech{\mathop{\rm sech}\nolimits}

\newcommand{\be}{\begin{equation}}
\newcommand{\ee}{\end{equation}}

\newcommand{\media}[1]{\langle #1 \rangle}

\begin{document}

\title{Exact results and new insights
for models defined over small world networks. \\
First and second order phase transitions. II: Applications}
\author{M. Ostilli }
\affiliation{Departamento de F{\'\i}sica da Universidade de Aveiro, 3810-193 Aveiro,
Portugal}
\affiliation{Center for Statistical Mechanics and Complexity, 
INFM-CNR SMC, Unit\`a di Roma 1, Roma, 00185, Italy.}
\author{J. F. F. Mendes}
\affiliation{Departamento de F{\'\i}sica da Universidade de Aveiro, 3810-193 Aveiro,
Portugal}

\begin{abstract}
We apply a novel method (presented in
part I) to solve several small-world models for which the method can be applied
analytically: the Viana-Bray model (which can be seen as a 0 or infinite
dimensional small-world model), the one-dimensional chain small-world
model, and the small-world spherical model in generic dimension.
In particular, we analyze in detail the one-dimensional chain small-world
model with negative short-range coupling showing that in this case,
besides a second-order spin glass phase transition, there
are two critical temperatures corresponding to first- or second-order
phase transitions.  
\end{abstract}

\maketitle

\email{ostilli@roma1.infn.it}

\section{Introduction} \label{intro}
This paper represents the continuation of another paper \cite{Part_I}
- in the following referred as part I - 
in which a general method to analyze small-world models has been
presented. Here (part II) we apply it to some cases
of interest which can be fully faced analytically.
Even if the exactness of this method is limited
to the paramagnetic regions, it gives the exact critical
behavior and the exact critical surfaces, 
and provide a clear and immediate (also in terms of calculation)
insight of the physics. 
As stressed in part I,
the underlying structure of the non random part of the model, \textit{i.e.},
the set of spins staying in a given lattice $\mathcal{L}_0$ of dimension $d_0$
and interacting through a fixed coupling $J_0$, is exactly
taken into account. When $J_0\geq 0$, the small-world effect gives rise
to the known fact that a second order phase transition takes place,
independently of the dimension $d_0$ and of the added random connectivity $c$.
However, when $J_0<0$, a completely different scenario emerges 
and - besides a second-order spin glass transition - for a sufficiently large $c$,
multiple first and second-order ferromagnetic phase transitions may take place.
Here we will emphasize above all this aspect which, 
to the best of our knowledge, until now was observed only 
in \cite{Bolle} and in the context of 
small-world neural networks in \cite{Skantos}, and in our opinion
represents a new paradigm in view of interesting applications
in real small-world networks.

The main advantage of our method lies in its great simplicity. 
In fact, what is required to apply it, is to solve not
the small-world model (a random model) defined over $\mathcal{L}_0$, 
but a corresponding non random model 
still defined over $\mathcal{L}_0$ and immersed in an external
magnetic field. This implies that we are able to solve analytically
small-world models whose underlying lattice 
$\mathcal{L}_0$ has dimension 0, 1 or infinite,
as for an ensemble of non interacting units, 
the one dimensional chain, and the spherical model, respectively.
In fact, in all these cases the non random model can be exactly solved even
when immersed in an external magnetic field. 

The paper is organized as follows.
In Secs. II and III we recall the definition of the small-world 
models and the method, which mainly consists in finding the solution
of the self-consistent equation (\ref{THEOa}) minimizing the effective
free energy (\ref{THEOll}). 
In Sec. IV we analyze two simple cases in which $\mathcal{L}_0$ has dimension
$d_0=0$: an ensemble of non interacting units. The small world-model
defined on it gives rise in particular to the well known Viana-Bray model
(sometime called random Bethe lattice model).
In Sec. V we consider the case in which $\mathcal{L}_0$ is the 
one dimensional chain. Sec. VI is devoted to the spherical model 
- an infinite-dimensional model - defined for arbitrary $d_0$ \cite{Note}.
Finally, in Sec. VII some conclusions are drawn.

\section{Small world models}
\label{models}
We consider random Ising models
constructed by super-imposing random graphs with finite average connectivity
onto some given lattice $\mathcal{L}_0$ whose set of bonds $(i,j)$ 
and dimension will be indicated by $\Gamma_0$ and $d_0$, respectively.
Given an Ising model - shortly \textit{the unperturbed model} - 
of $N$ spins coupled over $\mathcal{L}_0$
through a coupling $J_0$ and  
with Hamiltonian 
\begin{eqnarray}
H_0\equiv -J_{0}\sum_{(i,j)\in \Gamma_0}\sigma_{i}\sigma_{j}-h\sum_i \sigma_i
\label{H0},
\end{eqnarray}
and given an ensemble $\mathcal{C}$ 
of unconstrained random graphs $\bm{c}$, $\bm{c}\in\mathcal{C}$,
whose bonds are determined by the adjacency matrix elements $c_{i,j}=0,1$,
we define the corresponding small-world model 
- shortly \textit{the random model} -
as described by the following Hamiltonian
\begin{eqnarray}
\label{H}
H_{\bm{c};\bm{J}}\equiv H_0-\sum_{i<j} c_{ij}{J}_{ij}\sigma_{i}\sigma_{j},
\end{eqnarray}
the free energy $F$ and the averages $\overline{\media{\mathop{O}}^l}$
being defined in the usual (quenched) way as
\begin{eqnarray}
\label{logZ}
-\beta F\equiv \sum_{\bm{c}\in\mathcal{C}} P(\bm{c})\int d\mathcal{P}
\left(\{{J}_{i,j}\}\right)
\log\left(Z_{\bm{c};\bm{J}}\right),
\end{eqnarray} 
and 
\begin{eqnarray}
\label{O}
\overline{\media{\mathop{O}}^l}\equiv 
\sum_{\bm{c}\in\mathcal{C}} P(\bm{c}) \int d\mathcal{P}\left(\{{J}_{i,j}\}\right)
\media{\mathop{O}}^l, \quad l=1,2
\end{eqnarray} 

where $Z_{\bm{c};\bm{J}}$ 
is the partition function of the quenched system

\begin{eqnarray}
\label{Z}
Z_{\bm{c};\bm{J}}= \sum_{\{\sigma_{i}\}}
e^{-\beta H_{\bm{c};\bm{J}}\left(\{\sigma_i\}\}\right)}, 
\end{eqnarray} 

$\media{\mathop{O}}_{\bm{c};\bm{J}}$ the Boltzmann-average 
of the quenched system (note that $\media{\mathop{O}}_{\bm{c};\bm{J}}$ depends on the
given realization of the ${J}$'s and of $\bm{c}$:
$\media{\mathop{O}}=\media{\mathop{O}}_{\bm{c};\bm{J}}$;
for shortness we will often omit to write these dependencies)

\begin{eqnarray}
\label{OO}
\media{\mathop{O}}\equiv \frac{\sum_{\{\sigma_i\}}\mathop{O}_{\bm{c};\bm{J}}e^{-\beta 
H_{\bm{c};\bm{J}}\left(\{\sigma_i\}\right)}}{Z_{\bm{c};\bm{J}}}, 
\end{eqnarray} 

and $d\mathcal{P}\left(\{{J}_{i,j}\}\right)$ and $P(\bm{c})$ 
are two product measures given 
in terms of two normalized measures $d\mu(J_{i,j})\geq 0$ and $p(c_{i,j})\geq 0$, 
respectively: 
\begin{eqnarray}
\label{dP}
d\mathcal{P}\left(\{{J}_{i,j}\}\right)\equiv \prod_{(i,j),i<j} 
d\mu\left( {J}_{i,j} \right),
\quad \int d\mu\left( {J}_{i,j} \right)=1,
\end{eqnarray}
\begin{eqnarray}
\label{Pg}
P(\bm{c})\equiv \prod_{(i,j),i<j} p(c_{i,j}),
\quad \sum_{c_{i,j}=0,1} p(c_{i,j})=1.
\end{eqnarray}
The variables
$c_{i,j}\in\{0,1\}$ specify whether a ``long-range'' bond between the sites
$i$ and $j$ is present ($c_{i,j}=1$) or absent ($c_{i,j}=0$), whereas
the $J_{i,j}$'s are the random variables of the given bond $(i,j)$.
For the $c_{i,j}$'s, we shall consider the following 
distribution
\begin{eqnarray}
\label{PP}
 p(c_{ij})=
\frac{c}{N}\delta_{c_{ij},1}+\left(1-\frac{c}{N}\right)\delta_{c_{ij},0},
\end{eqnarray}
where $c>0$.
This choice leads in the thermodynamic limit $N\to\infty$ to a
number of long range connections per site distributed according
to a Poisson law with mean connectivity $c$.

When we will need to be specific, 
for the $J_{i,j}$'s we will assume either the distribution
\begin{eqnarray}
\label{dPF}
\frac{d\mu\left( {J}_{i,j} \right)}{d{J}_{i,j}}=\delta\left( {J}_{i,j}-J\right),
\end{eqnarray} 
or 
\begin{eqnarray}
\label{dPSG}
\frac{d\mu\left( {J}_{i,j} \right)}{d{J}_{i,j}}=
p\delta\left({J}_{i,j}-J\right)d{J}_{i,j}+
(1-p)\delta\left( {J}_{i,j}+J\right),
\end{eqnarray} 
to consider ferromagnetism or glassy phases, respectively.
In Eq. (\ref{dPSG}) $p\in [0,1]$.

\section{An effective field theory}
Depending on the temperature T, and 
on the parameters of the probability distributions, $d\mu$ and $p$,
the random model may stably stay either in the paramagnetic (P), 
in the ferromagnetic (F), or in the spin glass (SG) phase.
In our approach for the F and SG phases there are two natural order parameters
that will be indicated by $m^{(\mathrm{F})}$ and $m^{(\mathrm{SG})}$.
Similarly, for any correlation function, quadratic or not, there are
two natural quantities 
indicated by $C^{(\mathrm{F})}$ and $C^{(\mathrm{SG})}$, and that in turn
will be calculated in terms of $m^{(\mathrm{F})}$ and $m^{(\mathrm{SG})}$,
respectively. 
To avoid confusion, it should be kept in mind that
in our approach, for any
observable $\mathcal{O}$
there are - in principle - always 
two solutions that we label as F and SG,
but, for any temperature,
only one of the two solutions is stable and useful 
in the thermodynamic limit.

In the following, we will use the label $\mathop{}_0$ 
to specify that we are referring
to the unperturbed model with Hamiltonian (\ref{H0}).
Let $m_0(\beta J_0,\beta h)$ be the stable magnetization 
of the unperturbed model with coupling $J_0$ and in the presence of a uniform
external 
field $h$ at inverse temperature $\beta$. 
Then, the order parameters 
$m^{(\Sigma)}$, $\Sigma$=F,SG, 
satisfy the following self-consistent decoupled equations
\begin{eqnarray}
\label{THEOa}
m^{(\Sigma)}=m_0(\beta J_0^{(\Sigma)},
\beta J^{(\Sigma)}m^{(\Sigma)}+\beta h),
\end{eqnarray} 
where the effective couplings $J^{(\mathrm{F})}$, $J^{(\mathrm{SG})}$,
$J_0^{(\mathrm{F})}$ and $J_0^{(\mathrm{SG})}$ are given by
\begin{eqnarray}
\label{THEOb}
\beta J^{(\mathrm{F})}= c\int d\mu(J_{i,j})\tanh(\beta J_{i,j}),
\end{eqnarray} 
\begin{eqnarray}
\label{THEOc}
\beta J^{(\mathrm{SG})}= c\int d\mu(J_{i,j})\tanh^2(\beta J_{i,j}),
\end{eqnarray} 
\begin{eqnarray}
\label{THEOd}
J_0^{(\mathrm{F})}= J_0,
\end{eqnarray} 
and
\begin{eqnarray}
\label{THEOe}
\beta J_0^{(\mathrm{SG})}= \tanh^{-1}(\tanh^2(\beta J_0)).
\end{eqnarray}

For the correlation functions ${{C}}^{(\Sigma)}$, $\Sigma$=F,SG, we have
\begin{eqnarray}
\label{THEOh}
{{C}}^{(\Sigma)}=
{{C}}_0(\beta J_0^{(\Sigma)},\beta J^{(\Sigma)} m^{(\Sigma)}+\beta h)+\mathop{O}\left(\frac{1}{N}\right),
\end{eqnarray} 
where ${{C}}_0(\beta J_0,\beta h)$ is the correlation function of the unperturbed 
(non random) model.
For the corrective $\mathop{O}(1/N)$ term in Eq. (\ref{THEOh}) we remind the 
reader to Eq. (31) of part I.
Let us indicate by $C^{(\mathrm{1})}$ and $C^{(\mathrm{2})}$
the averages and the quadratic averages over the disorder 
of the correlation function of degree, say, $k$.
$C^{(\mathrm{1})}$ and $C^{(\mathrm{2})}$, are related to  
$C^{(\mathrm{F})}$ and $C^{(\mathrm{SG})}$, as follows
\begin{eqnarray}
\label{THEOa0}
C^{(\mathrm{1})}&=&C^{(\mathrm{F})}, \quad \mathrm{in~F}, \\
\label{THEOa01}
C^{(\mathrm{1})}&=& 0, \quad k ~ \mathrm{odd}, \quad \mathrm{in~SG}, \\
\label{THEOa02}
C^{(\mathrm{1})}&=&C^{(\mathrm{SG})}, \quad k ~ \mathrm{even}, \quad \mathrm{in~SG},
\end{eqnarray} 
and
\begin{eqnarray}
\label{THEOa03}
C^{(\mathrm{2})}&=&\left(C^{(\mathrm{F})}\right)^2, \quad \mathrm{in~F}, \\
\label{THEOa04}
C^{(\mathrm{2})}&=&\left(C^{(\mathrm{SG})}\right)^2, \quad \mathrm{in~SG}.
\end{eqnarray} 

Among all the possible stable 
solutions of Eqs. (\ref{THEOa}), in the thermodynamic
limit, for both $\Sigma$=F and $\Sigma$=SG, 
the true solution $\bar{m}^{(\Sigma)}$, or leading solution, 
is the one that minimizes $L^{(\Sigma)}$:
\begin{eqnarray}
\label{THEOlead}
L^{(\Sigma)}\left(\bar{m}^{(\Sigma)}\right)&=&
\min_{m\in [-1,1]} L^{(\Sigma)}\left(m\right),
\end{eqnarray} 
where 
\begin{eqnarray}
\label{THEOll}
L^{(\Sigma)}(m)\equiv 
\frac{\beta J^{(\Sigma)}\left(m\right)^2}{2}+
\beta f_0\left(\beta J_0^{(\Sigma)},\beta J^{(\Sigma)}m+\beta h\right),
\end{eqnarray} 
$f_0(\beta J_0,\beta h)$ being the free energy density in the thermodynamic
limit of the unperturbed model with coupling $J_0$ and in the presence
of an external field $h$, at inverse temperature $\beta$.
A necessary condition for a solution $m^{(\Sigma)}$ to be the leading solution
is the stability condition:
\begin{eqnarray}
\label{THEOgen}
{\tilde{\chi}_0\left(\beta^{(\Sigma)}J_0^{(\Sigma)},\beta J^{(\Sigma)}m^{(\Sigma)}+\beta h\right)}
\beta^{(\Sigma)}J^{(\Sigma)}<1,
\end{eqnarray} 
where $\tilde{\chi}_0(\beta J_0,\beta h)$ stands for the susceptibility $\chi_0$ of the
unperturbed model divided by $\beta$.

For the localization and the reciprocal stability between the F and SG phases 
we remind the reader to Sec. IIID of part I. We recall however that, 
at least for lattices $\mathcal{L}_0$
having only loops of even length, the stable P region is always that 
corresponding to a P-F phase diagram, so that in the P region
the correlation functions must be calculated only
through Eqs. (\ref{THEOa0}) and (\ref{THEOa03}).

If $J_0\geq 0$, the inverse critical temperature $\beta_c^{(\Sigma)}$
is solution of the following exact equation
\begin{eqnarray}
\label{THEOg}
{\tilde{\chi}_0\left(\beta_c^{(\Sigma)}J_0^{(\Sigma)},0\right)}
\beta_c^{(\Sigma)}J^{(\Sigma)}=1, \quad \beta_c^{(\Sigma)}<\beta_{c0}^{(\Sigma)},
\end{eqnarray} 
where $\beta_{c0}^{(\Sigma)}$
is the inverse critical temperature of the unperturbed model with
coupling $J_0^{(\Sigma)}$. The constrain in Eq. (\ref{THEOg}) ensures
the unicity of the solution. 

We end this section by stressing that this method is exact in
all the P region and provides the
exact critical surface, behavior and percolation threshold, and that,
in the absence of frustration, 
the order parameters $m^{(\Sigma)}$ become exact also
in the limit $c\to 0^+$, in the case of second-order phase transitions, 
and in the limit $c\to\infty$ (see Sec. IIIC of part I).
Note also that the order parameters $m^{(\Sigma)}$,
and then the correlation functions, are by
construction always exact in the zero temperature limit.

\section{Small World in $d_0=0$ Dimension}
\subsection{The Viana-Bray model}
As an immediate example, let us consider the Viana-Bray model.
It can be seen as the simplest small-world model in which
$N$ spins with no short-range couplings (here $J_0=0$) 
are randomly connected by long-range connections $J$ (possibly also random). 
Note that formally here $\mathcal{L}_0$ has dimension $d_0=0$.
Since $J_0=0$, for the unperturbed model we have
\begin{eqnarray}
\label{VBa}
-\beta f_0 (0,\beta h)=
\log\left[2\cosh(\beta h)\right],
\end{eqnarray} 
\begin{eqnarray}
\label{VBb}
m_0(0,\beta h)=\tanh(\beta h),
\end{eqnarray} 
\begin{eqnarray}
\label{VBc}
\tilde{\chi_0} (\beta J_0,\beta h)&=& 1-\tanh^2(\beta h)|_{\beta h=0}=1,
\end{eqnarray} 
It is interesting to check that the first and second derivatives of
$\tilde{\chi}_0$ in $h=0$ are null and negative, respectively. In fact we have
\begin{eqnarray}
\label{VBd}
&&\frac{\partial}{\beta h}\tilde{\chi_0} (0,\beta h)=
-2\tanh^2(\beta h)
\nonumber \\
&& \times \left[1-\tanh^2(\beta h)\right]|_{\beta h=0}=0,
\end{eqnarray} 
and
\begin{eqnarray}
\label{VBe}
&& \frac{\partial^2}{(\beta h)^2}\tilde{\chi_0} (0,\beta h)=
-2\left[1-\tanh^2(\beta h)\right]^2 +\nonumber \\
&& 4\tanh^2(\beta h)\left[1-\tanh^2(\beta h)\right]
|_{\beta h=0}=-2.
\end{eqnarray} 

Applying these results to Eqs. (\ref{THEOa}-\ref{THEOe}) we get immediately
the self-consistent equations for the F and the SG magnetizations
\begin{eqnarray}
\label{VBf}
m^{(\mathrm{F})}&=&\tanh\left[m^{(\mathrm{F})}c\int d\mu\tanh(\beta J)\right],\\
\label{VBg}
m^{(\mathrm{SG})}&=&\tanh\left[m^{(\mathrm{SG})}c\int d\mu\tanh^2(\beta J)\right], 
\end{eqnarray} 
and the Viana-Bray critical surface
\begin{eqnarray}
\label{VBh}
c\int d\mu\tanh(\beta_c^{(\mathrm{F})} J)=1,\\
\label{VBi}
c\int d\mu\tanh^2(\beta_c^{(\mathrm{SG})} J)=1.
\end{eqnarray} 

On choosing for $d\mu$ a measure having average 
and variance scaling as $\mathop{O}(1/c)$,
for $c\propto N$, we recover
the equations for the Sherrington-Kirkpatrick model 
already derived in this form in \cite{MOI} and \cite{MOII}. 
In these papers Eqs. (\ref{VBf}-\ref{VBi}) were derived by mapping the
Viana-Bray model and, similarly, the Sherrington-Kirkpatrick model to
the non random fully connected Ising model. In this sense it should be also clear
that, at least for $\beta\leq \beta_c$ and zero external field, 
in the thermodynamic limit, 
the connected correlation functions (of order $k$ greater than 1) 
in the Sherrington-Kirkpatrick 
and in the Viana-Bray model are exactly zero. In fact, in the thermodynamic
limit, the non random fully connected model can be exactly reduced to a model 
of non interacting spins immersed in
an effective medium so that among any two spins there is no correlation.
Such a result is due to the fact that, in these models, 
all the $N$ spins interact 
through the same coupling $J/N$, no matter how far apart they are, and the net effect
of this is that in the thermodynamic limit the system becomes equivalent to
a collection of $N$ non interacting spins seeing only an effective external
field (the medium) like in Eq. (\ref{VBb}) with $\beta h \to \beta J m_0$.

In the limit $\beta \to \infty$, Eqs. (\ref{VBf}) and (\ref{VBg})
give the following size (normalized to 1) of the giant connected component
\begin{eqnarray}
\label{VBff}
m^{(\mathrm{F})}&=&\tanh(m^{(\mathrm{F})}c),\\
\label{VBgg}
m^{(\mathrm{SG})}&=&\tanh(m^{(\mathrm{SG})}c). 
\end{eqnarray} 

These equations are not exact, however they succeed in giving the 
exact percolation threshold $c=1$. In fact, concerning the equation (\ref{VBff})
for the F phase, the exact equation for $m^{(\mathrm{F})}$ is (see for example 
\cite{Review} and references therein)
\begin{eqnarray}
\label{VBff1}
1-m^{(\mathrm{F})}&=&e^{-cm^{(\mathrm{F})}},
\end{eqnarray} 
which, in terms of the function $\tanh$, becomes
\begin{eqnarray}
\label{VBff2}
\frac{2m^{(\mathrm{F})}+(m^{(\mathrm{F})})^2}{2-m^{(\mathrm{F})}+(m^{(\mathrm{F})})^2}
&=&\tanh(m^{(\mathrm{F})}c),
\end{eqnarray} 
so that Eqs. (\ref{VBff}) and (\ref{VBff1}) are equivalent at the order 
$\mathop{O}(m^{(\mathrm{F})})$. We see also that, as stated in the previous Section,
Eqs. (\ref{VBff}) and (\ref{VBff1}) become equal in the limits $c\to0$ and $c\to\infty$.

\subsection{Gas of Dimers}
Let us consider for $\mathcal{L}_0$ a set of $2N$ spins coupled through a coupling $J_0$
two by two. The expression ``gas of dimers'' stresses the fact that the dimers,
\textit{i.e.} the couples of coupled spins, do not interact each other.
As a consequence, the free energy,
the magnetization, and the susceptibility of the unperturbed model
can be immediately calculated. We have
\begin{eqnarray}
\label{gd}
-\beta f_0 (\beta J_0,\beta h)=\frac{1}{2}
\log\left[2e^{\beta J_0}\cosh(2\beta h)+2e^{-\beta J_0}\right],
\end{eqnarray} 
\begin{eqnarray}
\label{gd1}
m_0 (\beta J_0,\beta h)=\frac{e^{\beta J_0}\sinh(2\beta h)}
{e^{\beta J_0}\cosh(2\beta h)+e^{-\beta J_0}},
\end{eqnarray} 
\begin{eqnarray}
\label{gd1}
\tilde{\chi_0} (\beta J_0,\beta h)&=&\frac{2e^{\beta J_0}+2\cosh(2\beta h)}
{\left[e^{\beta J_0}\cosh(2\beta h)+e^{-\beta J_0}\right]^2}|_{\beta h=0}\nonumber \\
&=& \frac{e^{\beta J_0}}{\cosh(\beta J_0)},
\end{eqnarray} 
Let us calculate also the second derivative of $\tilde{\chi}_0$. From
\begin{eqnarray}
\label{gd2}
&&\frac{\partial}{\beta h}\tilde{\chi_0} (\beta J_0,\beta h)=
4\sinh(\beta h) \nonumber \\
&\times& \frac{e^{-\beta J_0}-2e^{3\beta J_0}
-e^{\beta J_0}\cosh(2\beta h)}
{\left[e^{\beta J_0}\cosh(2\beta h)+e^{-\beta J_0}\right]^3},
\end{eqnarray} 
we get
\begin{eqnarray}
\label{gd3}
\frac{\partial^2}{(\beta h)^2}\tilde{\chi_0} (\beta J_0,\beta h)|_{\beta h=0}=
-2\frac{\sinh(\beta J_0)+e^{3\beta J_0}}{\left[\cosh(\beta J_0)\right]^3}.
\end{eqnarray} 
We note that, as expected, the second derivative of $\tilde{\chi}_0$ in $h=0$,
for $J_0\geq 0$ is always negative, whereas, for $J_0<0$ it becomes positive as soon as 
$\beta |J_0|>\log(\sqrt{2})$. 

By using the above equations, 
from Sec. III we get immediately the following self-consistent equation for the
magnetizations 
\begin{eqnarray}
\label{gd4}
m^{(\Sigma)}=\frac{\tanh(2\beta J^{(\Sigma)}m^{(\Sigma)}+ 2\beta h)}
{1+e^{-2\beta J_0}\sech(2\beta J^{(\Sigma)}m^{(\Sigma)}+ 2\beta h)},
\end{eqnarray} 
and - at least for $J_0\geq 0$ - the equation for the critical temperature 
\begin{eqnarray}
\label{gd5}
\frac{e^{\beta_c^{(\Sigma)}J_0^{(\Sigma)}}}
{\cosh(\beta_c^{(\Sigma)} J_0^{(\Sigma)})}
\beta_c^{(\Sigma)} J^{(\Sigma)}=1.
\end{eqnarray} 

As it will be clear soon, this model lies between the Viana-Bray 
model and the more complex $d_0=1$ dimensional chain small-world model, which will
be analyzed in detail in the next section.
Our major interest in this simpler gas of dimer small-world model 
is related to the fact that, 
in spite of its simplicity and $d_0$=0 dimensionality - since 
the second derivative of $\tilde{\chi}_0$ may be positive when $J_0$ is
negative -
according to the general result of Sec. IIIB of part I - 
it is already able to give rise to 
also multiple first- and second-order phase transitions. 

\section{Small world in $d_0$=1 dimension}
In this section we will analyze the case in which $\mathcal{L}_0$ is the 
$d_0$=1-dimensional chain with periodic boundary conditions (p.b.c.). 
The corresponding small-world model with Hamiltonian (\ref{H}) in zero field
has already been analyzed in \cite{Niko} by using the replica method. Here we will
recover the results found in \cite{Niko} for $\beta_c$ and will provide the
self-consistent equations for the magnetizations $m^{(\mathrm{F})}$ and
$m^{(\mathrm{SG})}$  whose solution, as expected,
turns out to be in good agreement with the corresponding solutions found in
\cite{Niko} for $c$ small. It will be however rather evident how much the two methods 
differ in terms of simplicity and intuitive meaning. 
Furthermore, we will derive also an explicit expression for the two-points
connected correlation function which, to the best of our knowledge, had not been
published yet. Finally, we will analyze in the detail the completely novel scenario
for the case $J_0<0$ which, as mentioned, produces
multiple first- and second-order phase transitions.

In order to apply the method of Sec.III we have to solve the one dimensional
Ising model with p.b.c. immersed in an external field. 
The solution of this non random model is easy and 
well known (see for example \cite{Baxter}). 
For the free energy density, the magnetization 
and the two-points connected correlation function we have
\begin{eqnarray}
\label{1d}
-\beta f_0 (\beta J_0,\beta h)=\log\left( \lambda_1\right),
\end{eqnarray} 

\begin{eqnarray}
\label{1da}
m_0 (\beta J_0,\beta h)=\frac{e^{\beta J_0}\sinh(\beta h)}
{\left[e^{2\beta J_0}\sinh^2(\beta h)+e^{-2\beta J_0}\right]^{\frac{1}{2}}},
\end{eqnarray} 
\begin{eqnarray}
\label{1db}
{{C}}_0 (\beta J_0,\beta h;{||i-j||}_0)&\equiv&{\media{\sigma_i\sigma_j}}_0-{\media{\sigma}}_0^2
\nonumber \\
 &=& \sin^2(2\varphi) \left(\frac{\lambda_2}{\lambda_1}\right)^{{||i-j||}_0},
\end{eqnarray} 
where the factor $\varphi$ is defined by  
\begin{eqnarray}
\label{1dc}
\cot(2\varphi)=e^{2\beta J_0}\sinh(\beta h), \quad 0<\varphi<\frac{\pi}{2},
\end{eqnarray} 
%
${||i-j||}_0$ is the (euclidean) distance between $i$ and $j$,
and $\lambda_1$ and $\lambda_2$ are the two greatest eigenvalues of the matrix
appearing in the transfer matrix method, whose ratio is given by
(in the following expression 
the numerator and the denominator are $\lambda_2$ and $\lambda_1$, respectively)
\begin{eqnarray}
\label{1dd}
\frac{\lambda_2}{\lambda_1}=
\frac
{e^{\beta J_0}\cosh(\beta h)-
\left[e^{2\beta J_0}\sinh^2(\beta h)+e^{-2\beta J_0}\right]^{\frac{1}{2}}}
{e^{\beta J_0}\cosh(\beta h)+
\left[e^{2\beta J_0}\sinh^2(\beta h)+e^{-2\beta J_0}\right]^{\frac{1}{2}}}.
\end{eqnarray} 

Let us calculate $\tilde{\chi}_0$ and its first and second derivatives.
From Eq. (\ref{1da}) we have
\begin{eqnarray}
\label{1de}
\tilde{\chi}_0 (\beta J_0,\beta h)=\frac{e^{-\beta J_0}\cosh(\beta h)}
{\left[e^{2\beta J_0}\sinh^2(\beta h)+e^{-2\beta J_0}\right]^{\frac{3}{2}}},
\end{eqnarray} 

\begin{eqnarray}
\label{1df}
&&\frac{\partial}{\partial \beta h}\tilde{\chi}_0 (\beta J_0,\beta h)=\sinh(\beta h)
\nonumber \\ &\times&
\frac{\left[e^{-3\beta J_0}-2e^{\beta J_0}\cosh^2(\beta h)-e^{\beta J_0}\right]}
{\left[e^{2\beta J_0}\sinh^2(\beta h)+e^{-2\beta J_0}\right]^{\frac{5}{2}}},
\end{eqnarray}
 
\begin{eqnarray}
\label{1dg}
&& \frac{\partial^2}{\partial (\beta h)^2}\tilde{\chi}_0 (\beta J_0,\beta h)=\cosh(\beta h)
\nonumber \\ &\times&
\frac{\left[e^{-3\beta J_0}-2e^{\beta J_0}\cosh^2(\beta h)-e^{\beta J_0}\right]}
{\left[e^{2\beta J_0}\sinh^2(\beta h)+e^{-2\beta J_0}\right]^{\frac{5}{2}}}
+\mathop{O}(\beta h)^2.
\end{eqnarray} 

From Eq. (\ref{1dg}) we see that for $J_0>0$ and any $\beta$ we have,
for sufficiently small $h$,
\begin{eqnarray}
\label{1dh}
&& \frac{\partial^2}{\partial (\beta h)^2}\tilde{\chi}_0 (\beta J_0,\beta h)<\cosh(\beta h)
\nonumber \\ &\times&
\frac{\left[1-3e^{\beta J_0}\right]}
{\left[e^{2\beta J_0}\sinh^2(\beta h)+e^{-2\beta J_0}\right]^{\frac{5}{2}}}
+\mathop{O}(\beta h)^2<0, 
\end{eqnarray} 
whereas for $J_0<0$ we have
\begin{eqnarray}
\label{1di}
&& \frac{\partial^2}{\partial (\beta h)^2}\tilde{\chi}_0 (\beta J_0,\beta h) \geq 0
\quad \mathrm{for} \quad e^{-4\beta J_0}>3.
\end{eqnarray} 

We see therefore that, according to Sec. IIIB of part I,
when $J_0<0$ for $\beta |J_0|\geq \log(3)/4=0.1193...$
we have a first-order phase transition.

From Eqs. (\ref{THEOa}) and (\ref{1da}), for the magnetizations $m^{(\mathrm{F})}$ and
$m^{(\mathrm{SG})}$ at zero external field we have
\begin{eqnarray}
\label{1dl}
m^{(\Sigma)}=\frac{e^{\beta J_0^{(\Sigma)}}\sinh(\beta J^{(\Sigma)}m^{(\Sigma)})}
{\left[e^{2\beta J_0^{(\Sigma)}}\sinh^2(\beta J^{(\Sigma)} 
m^{(\Sigma)})+e^{-2\beta J_0^{(\Sigma)}}\right]^{\frac{1}{2}}}.
\end{eqnarray} 

From Eqs. (\ref{THEOg}) and (\ref{1de}) we see that a solution $m^{(\Sigma)}$
becomes unstable at the inverse temperature $\beta_c^{(\Sigma)}$ given by
\begin{eqnarray}
\label{1dm}
e^{2\beta_c^{(\Sigma)} J_0^{(\Sigma)}}\beta_c^{(\Sigma)}J^{(\Sigma)}=1.
\end{eqnarray} 
For $J_0\geq 0$ the above equation gives 
the exact P-F and P-SG critical temperatures in agreement with \cite{Niko}.
When $J_0<0$ - unless the transition be of second-order - Eq. (\ref{1dm}) for $\Sigma=$F
does not signal a phase transition. In general, as $J_0<0$ the P-F critical temperature
must be determined by looking at all the stable solutions $m$ of the self-consistent 
equation (\ref{1dl}) and by choosing the one
minimizing the effective free energy $L^{\mathrm{(F)}}(m)$ of Eq. (\ref{THEOll}).

Finally, for the two-point connected correlation function,
from Eqs. (\ref{THEOh}), (\ref{1db}) and (\ref{1dc}), we have
\begin{eqnarray}
\label{1dp}
{{C}}^{(\Sigma)}({||i-j||}_0)&=& 
\sin^2(2\varphi^{(\Sigma)}) e^{-{||i-j||}_0/{\xi^{(\Sigma)}}}, 
\end{eqnarray} 
where
\begin{eqnarray}
\label{1dq}
2\varphi^{(\Sigma)}= \cot^{-1} 
\left[e^{2\beta J_0^{(\Sigma)}}
\sinh(\beta J^{(\Sigma)} m^{(\Sigma)})\right], 
\end{eqnarray} 
and the correlation length $\xi^{(\Sigma)}$ is given by performing 
the effective substitutions $\beta J_0\to \beta J_0^{(\Sigma)}$ and
$\beta h\to \beta J^{(\Sigma)} m^{(\Sigma)}$ in the logarithm of Eq. (\ref{1dd}).

Note that ${{C}}_0 (\beta J_0,\beta h)$ is even in $\beta h$, so that $C(-m)=C(m)$.

Near the critical temperature we have
\begin{eqnarray}
\label{1dr}
\sin(2\varphi^{(\Sigma)})=1-\frac{\left(e^{2\beta J_0^{(\Sigma)}}m^{(\Sigma)}\right)^2}{2}
+\mathop{O}(m^{(\Sigma)})^4
\end{eqnarray} 
and 
\begin{widetext}
\begin{eqnarray}
\label{1ds}
\left(\xi^{(\Sigma)}\right)^{-1}=\left|\log[\tanh(\beta J_0^{(\Sigma)})]-
\frac{(\beta J^{(\Sigma)} m^{(\Sigma)})^2}{4\sinh(\beta J_0^{(\Sigma)})}
\left[\left(e^{\beta J_0^{(\Sigma)}}+e^{3\beta J_0^{(\Sigma)}}\right)
\tanh(\beta J_0^{(\Sigma)})
+e^{3\beta J_0^{(\Sigma)}}-e^{\beta J_0^{(\Sigma)}}
\right] +\mathop{O}(m^{(\Sigma)})^4\right| .
\end{eqnarray} 
\end{widetext}
According to the general result (see Eqs. (51)-(55) of part I),
we see that the correlation length remains finite at all temperatures.

\begin{figure}
\epsfxsize=65mm \centerline{\epsffile{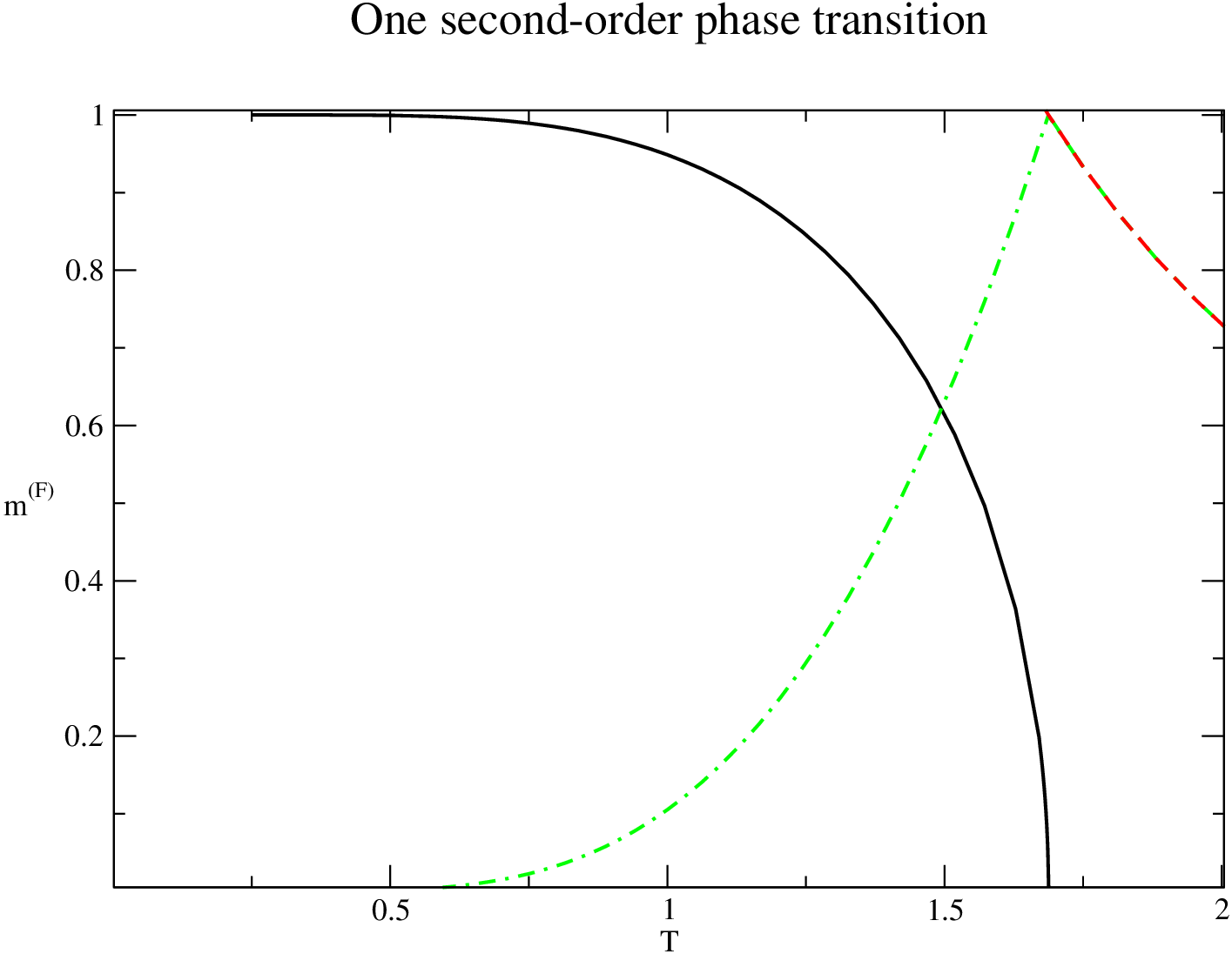}}
\caption{Magnetization (solid line), and curves of stability 
(dashed and dot-dashed lines) for the case
$c=0.5$, $J_0=1$ and $J=3/5/$. Here $T_{c}=1.687$.} 
\label{sw_1_pos}
\end{figure}
\begin{figure}
\epsfxsize=65mm \centerline{\epsffile{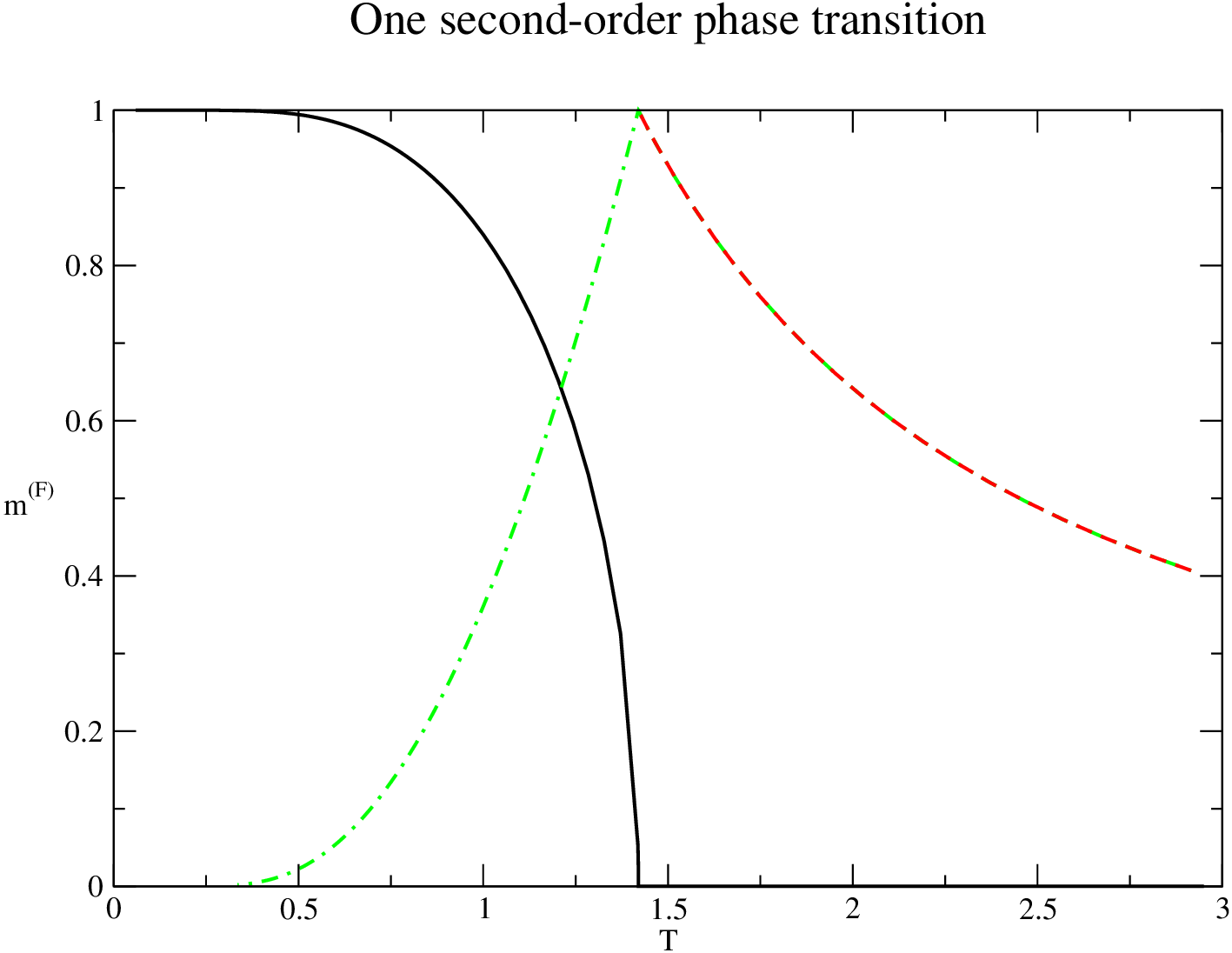}}
\caption{Magnetization (solid line), and curves of stability 
(dashed and dot-dashed lines) for the case
$c=10$, $J_0=0.25$ and $J=1/c$. Here $T_{c}=1.419$.} 
\label{sw_2_pos}
\end{figure}
\begin{figure}
\epsfxsize=65mm \centerline{\epsffile{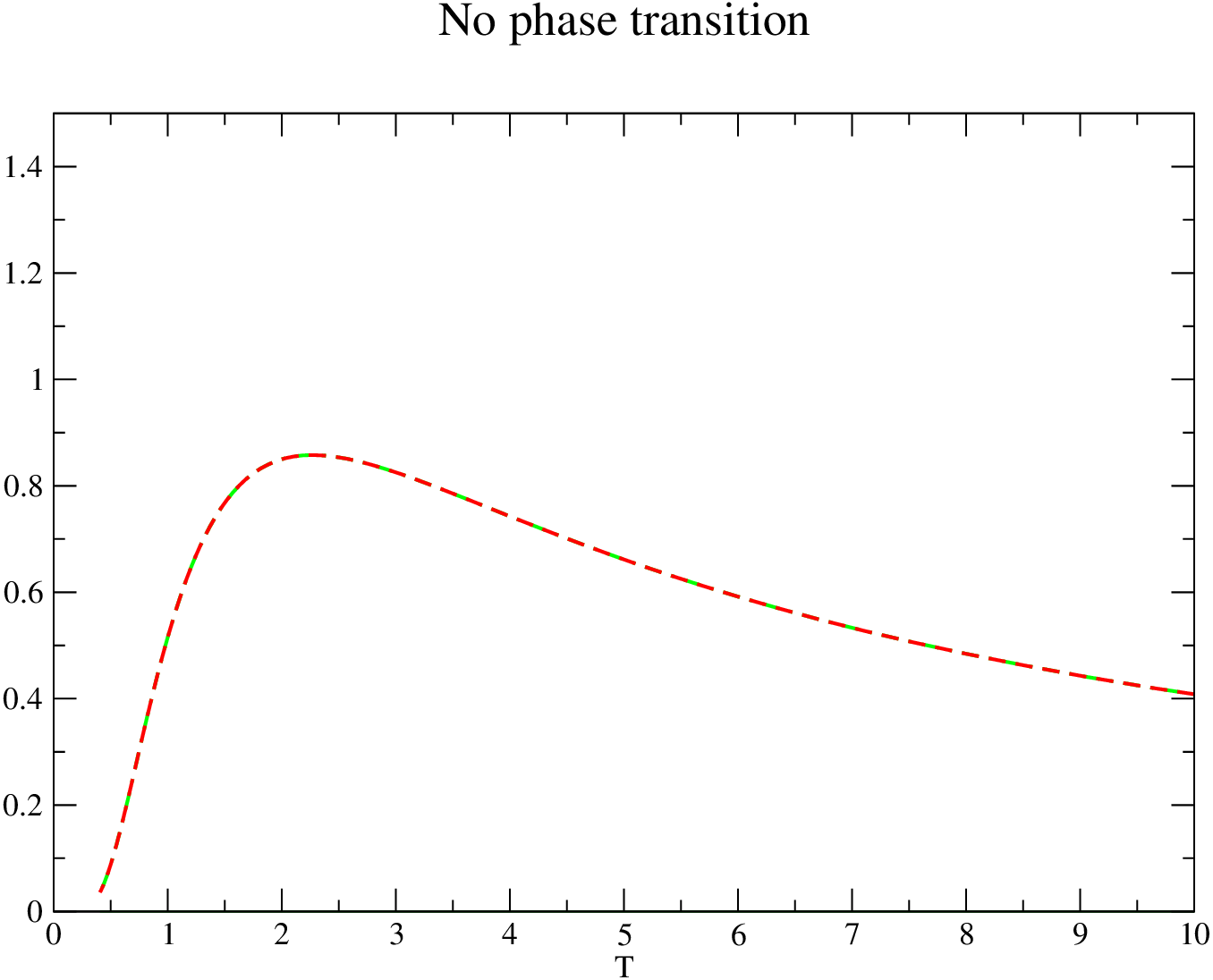}}
\caption{Curves of stability 
(dashed and dot-dashed lines) for the case
$c=5$, $J_0=-1$ and $J=1$.} 
\label{sw_1}
\end{figure}
\begin{figure}
\epsfxsize=65mm \centerline{\epsffile{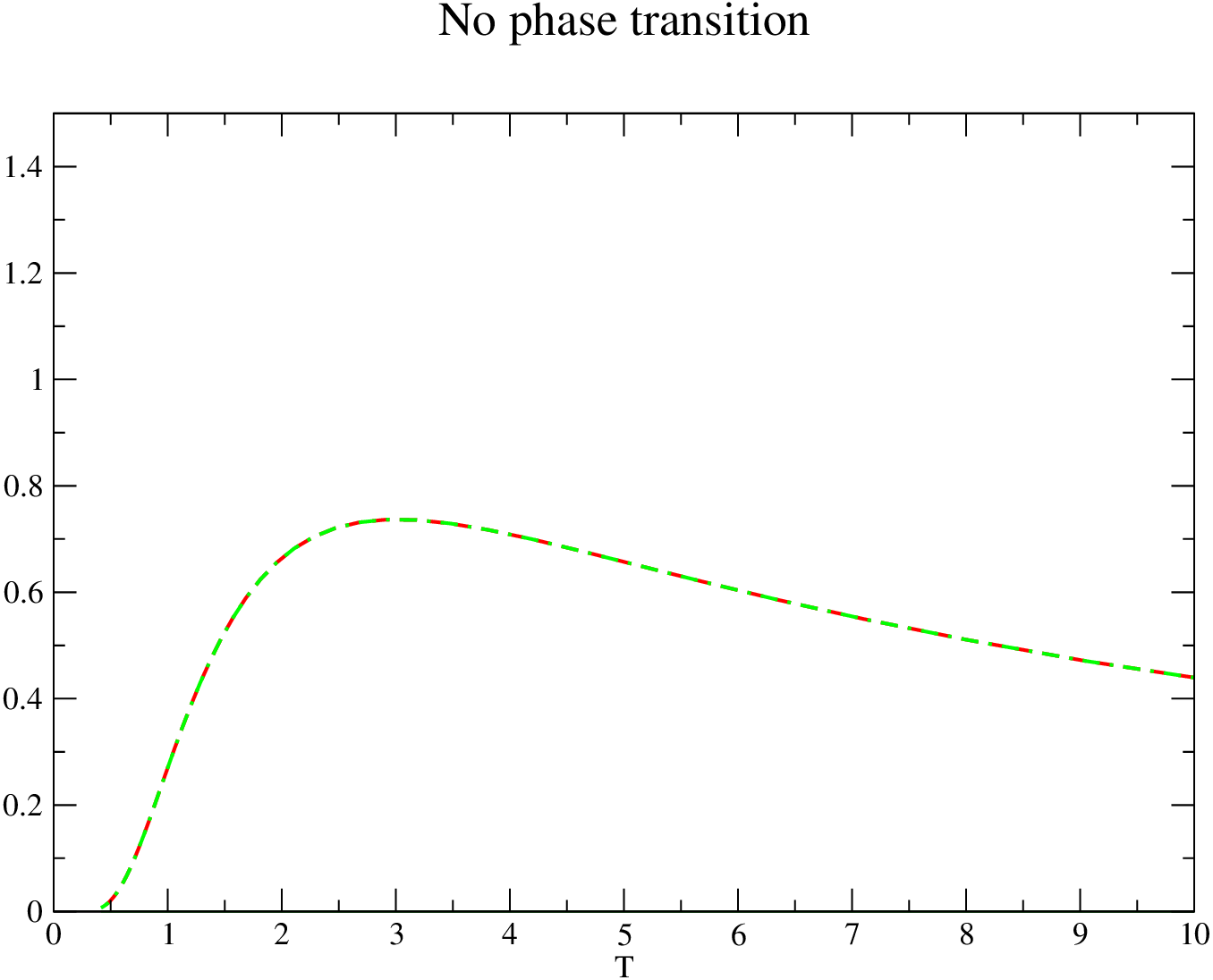}}
\caption{Curves of stability 
(dashed and dot-dashed lines) for the case
$c=5.828$, $J_0=-1.4$ and $J=1$.} 
\label{sw_2}
\end{figure}
\begin{figure}
\epsfxsize=65mm \centerline{\epsffile{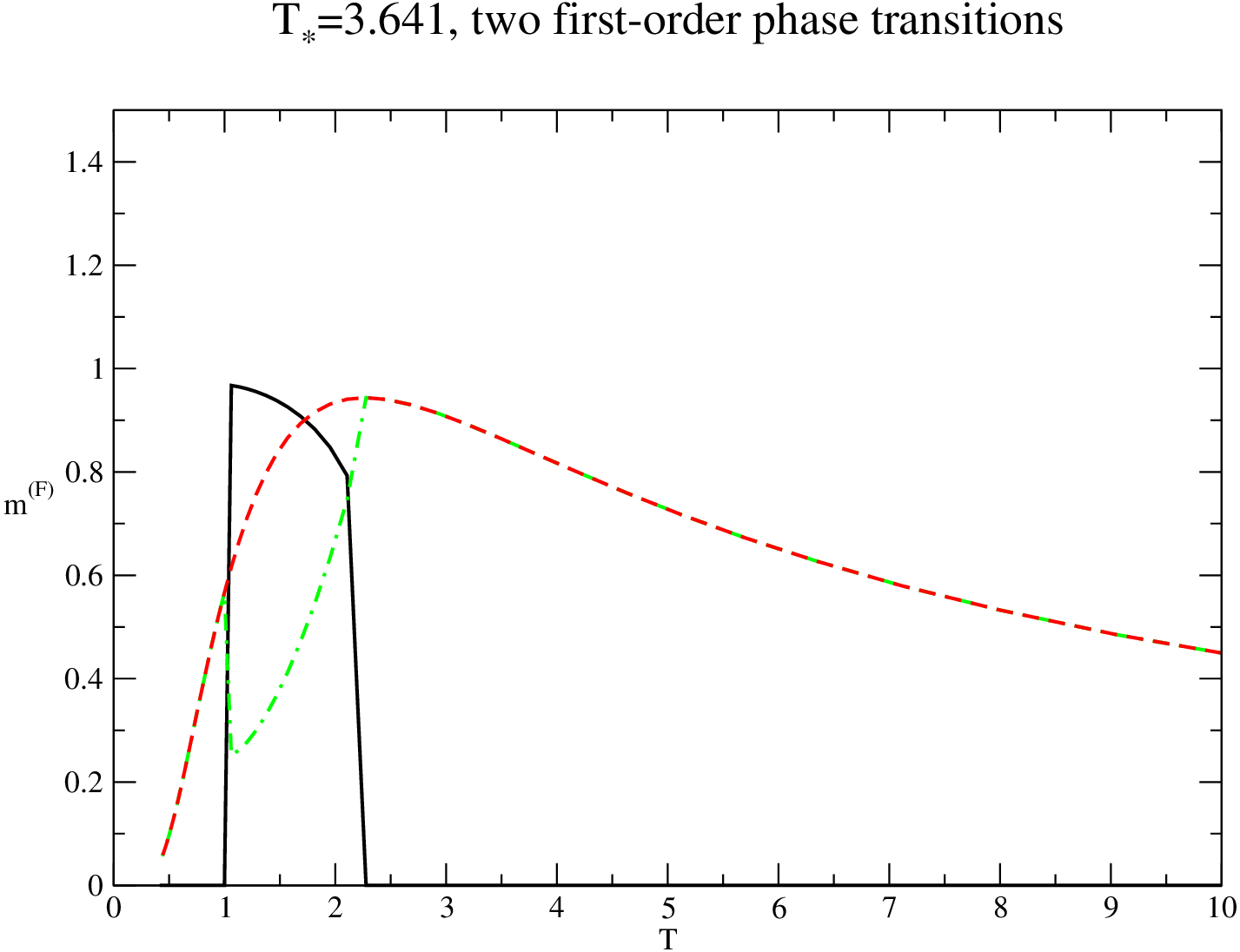}}
\caption{Magnetization (solid line), and curves of stability 
(dashed and dot-dashed lines) for the case
$c=5.5$, $J_0=-1$, and $J=1$. Here $T_{c1}=1.02$ and $T_{c2}=2.27$.} 
\label{sw_3}
\end{figure}
\begin{figure}
\epsfxsize=65mm \centerline{\epsffile{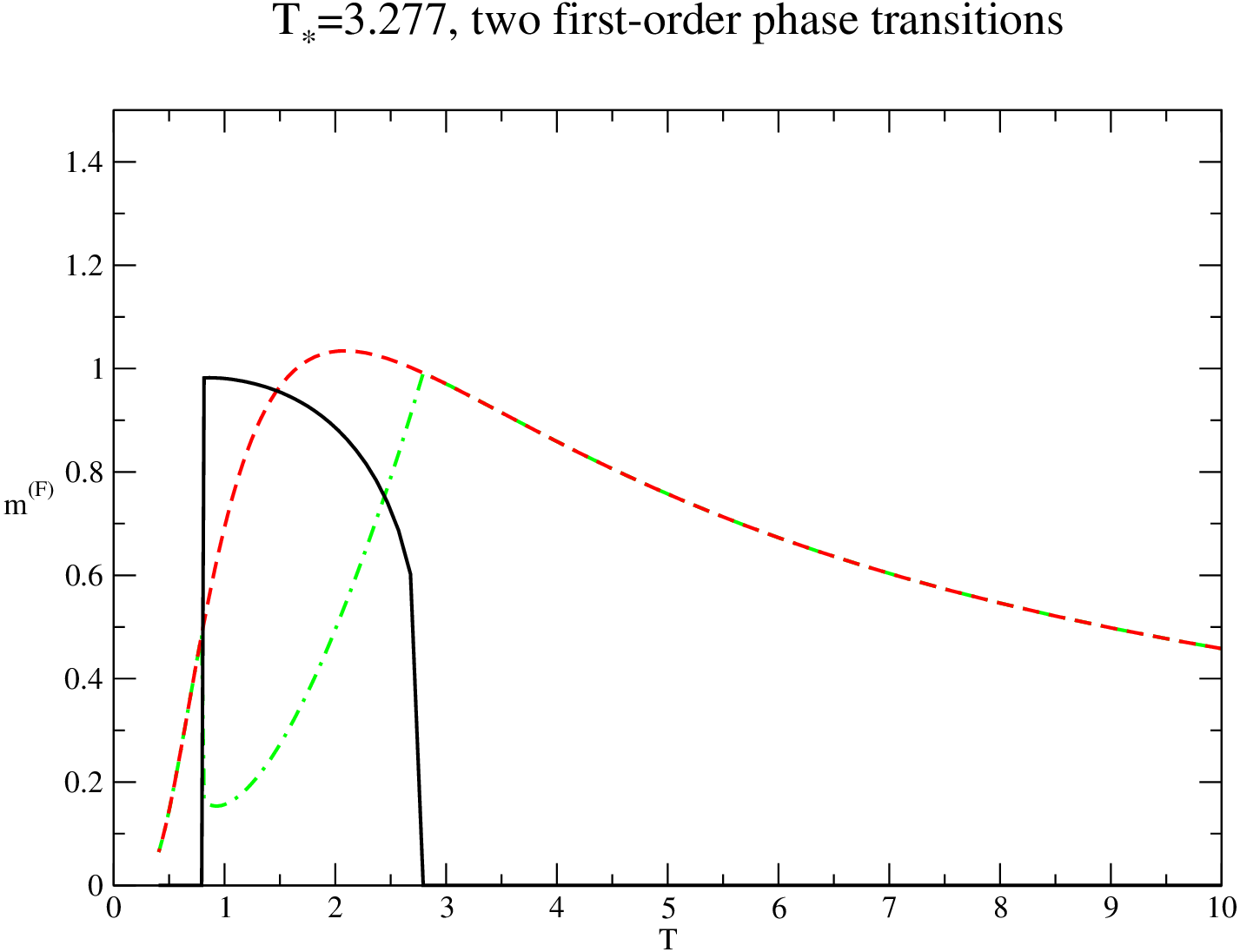}}
\caption{Magnetization (solid line), and curves of stability 
(dashed and dot-dashed lines) for the case
$c=5.5$, $J_0=-0.9$ and $J=1$. Here $T_{c1}=0.85$ and $T_{c2}=2.81$.} 
\label{sw_4}
\end{figure}
\begin{figure}
\epsfxsize=65mm \centerline{\epsffile{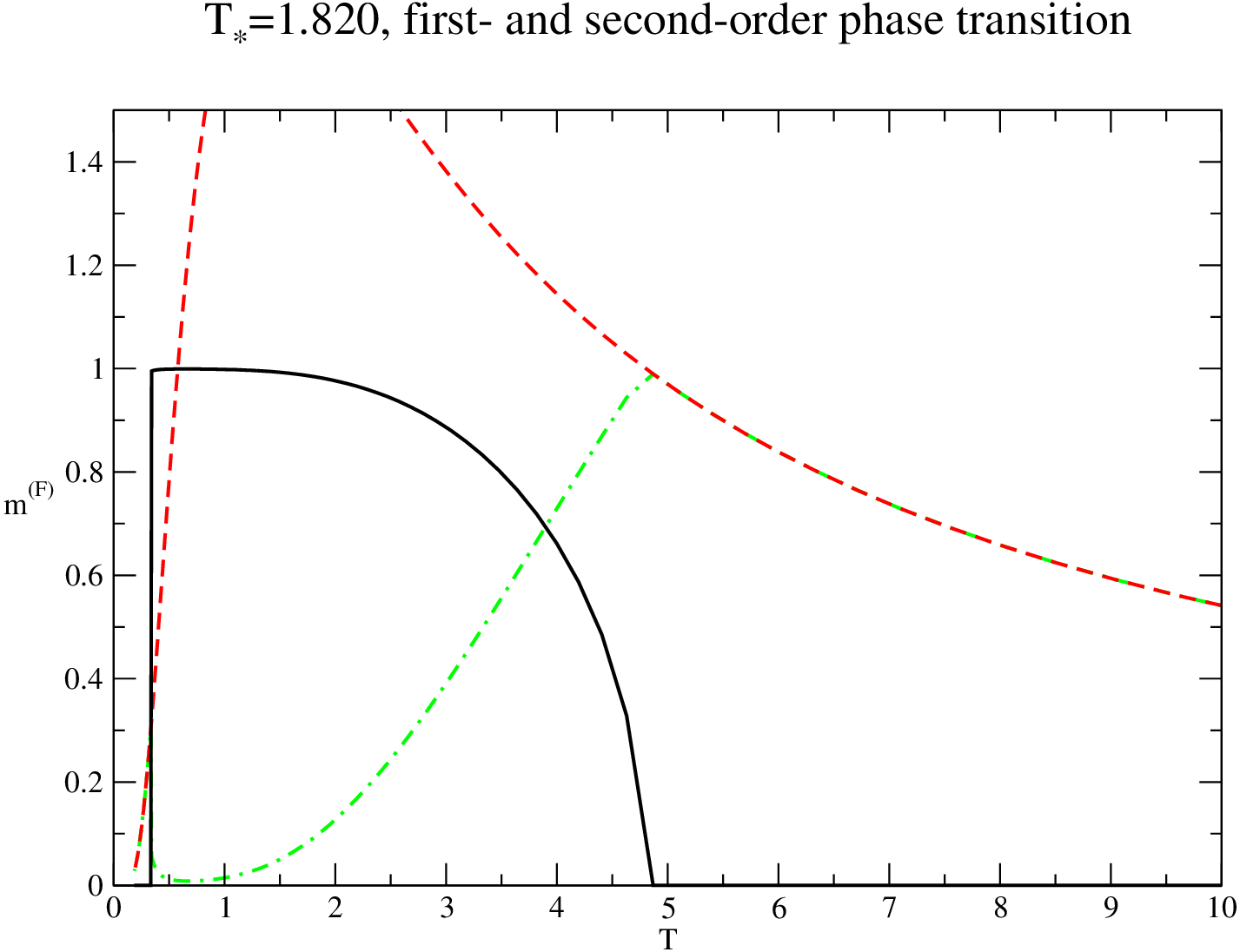}}
\caption{Magnetization (solid line), and curves of stability 
(dashed and dot-dashed lines) for the case
$c=6$, $J_0=-0.5$ and $J=1$. Here $T_{c1}=0.35$ and $T_{c2}=4.87$.
$T_{c2}$ corresponds to a second-order phase transition.} 
\label{sw_5}
\end{figure}
\begin{figure}
\epsfxsize=65mm \centerline{\epsffile{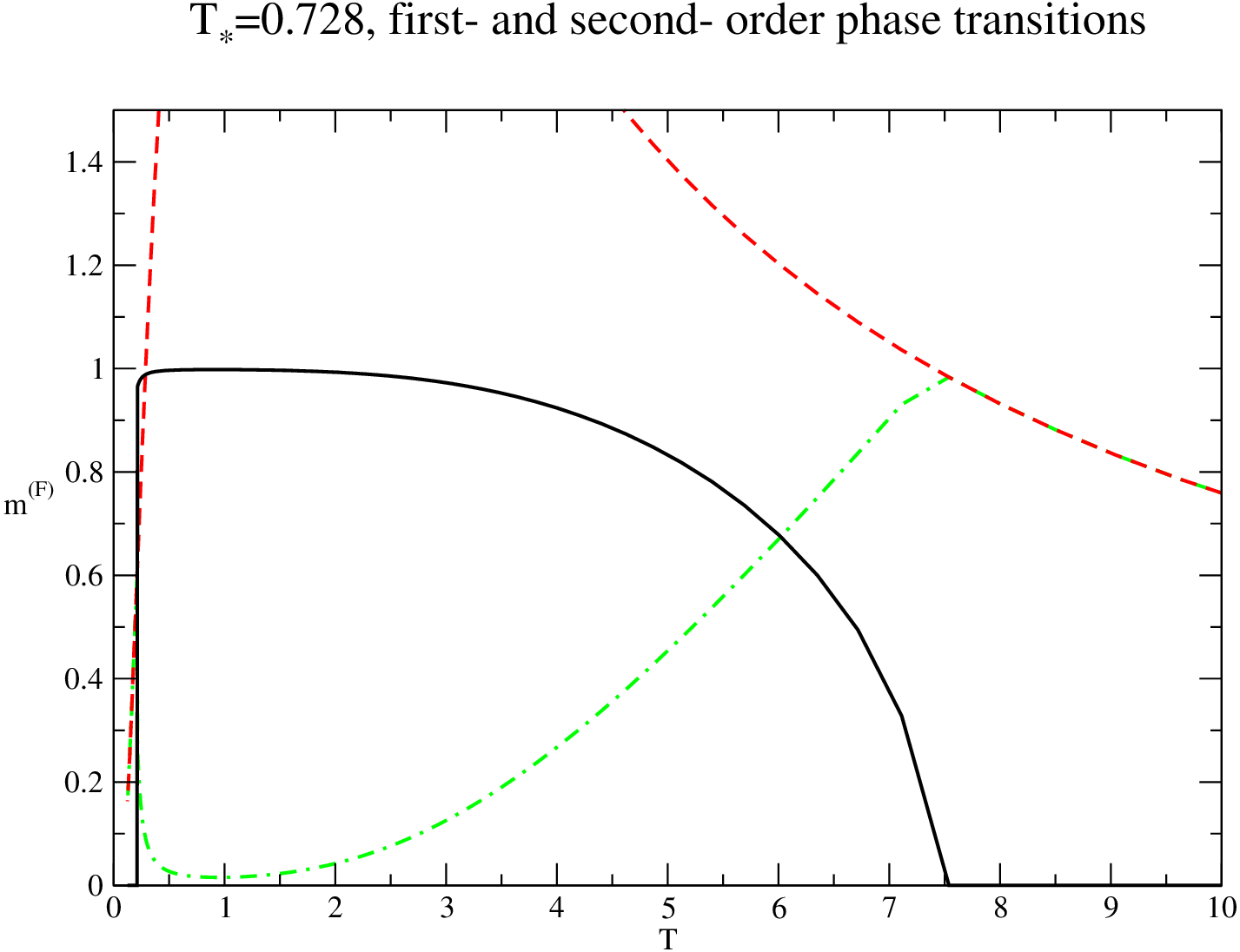}}
\caption{Magnetization (solid line), and curves of stability 
(dashed and dot-dashed lines) for the case
$c=4$, $J_0=-0.2$ and $J=2$. Here $T_{c1}=0.22$ and $T_{c2}=7.55$.
$T_{c2}$ corresponds to a second-order phase transition.} 
\label{sw_6}
\end{figure}
\begin{figure}
\epsfxsize=65mm \centerline{\epsffile{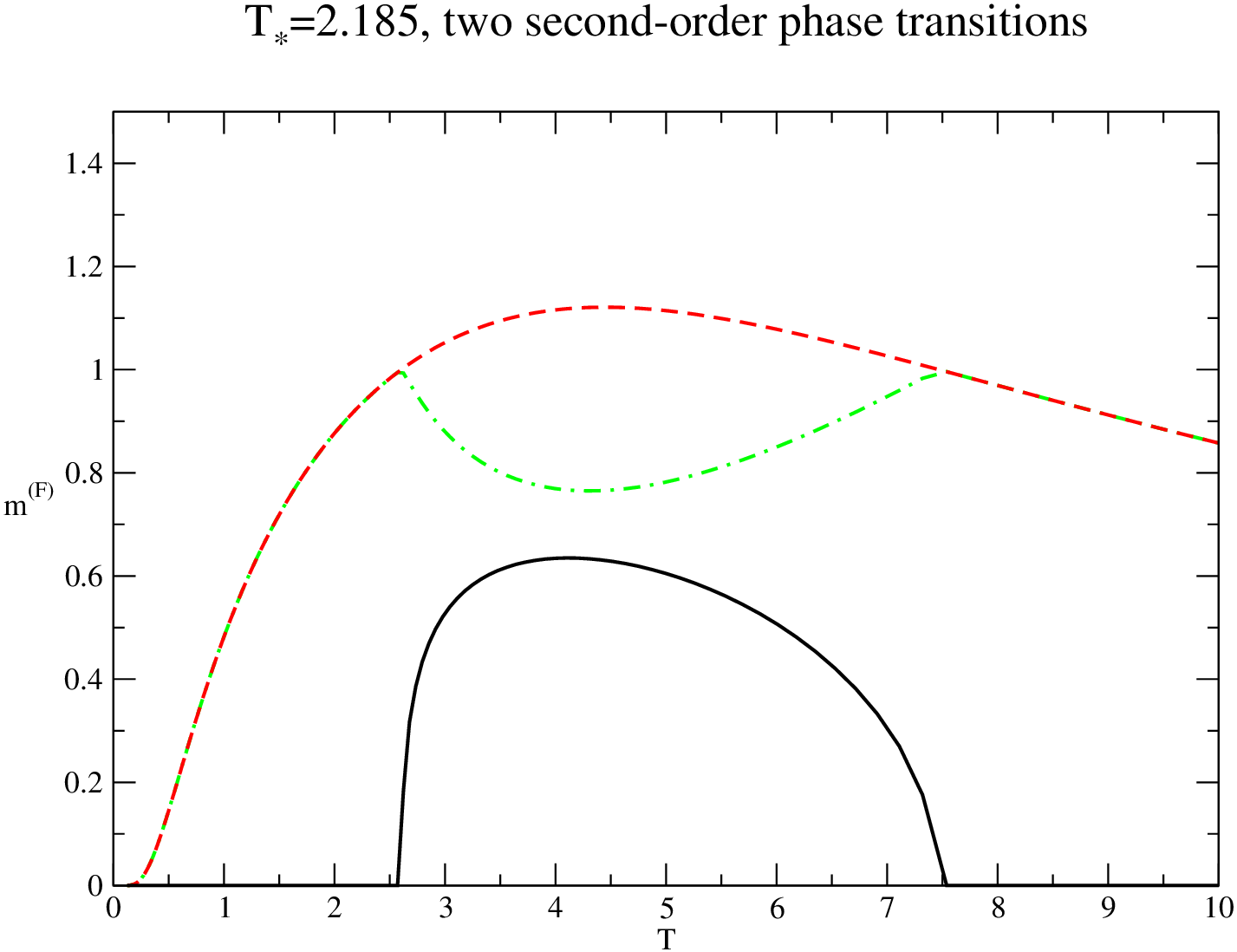}}
\caption{Magnetization (solid line), and curves of stability 
(dashed and dot-dashed lines) for the case
$c=1.6$, $J_0=-0.6$ and $J=7$. Here $T_{c1}=2.58$ and $T_{c2}=7.55$.} 
\label{sw_7}
\end{figure}
\begin{figure}
\epsfxsize=65mm \centerline{\epsffile{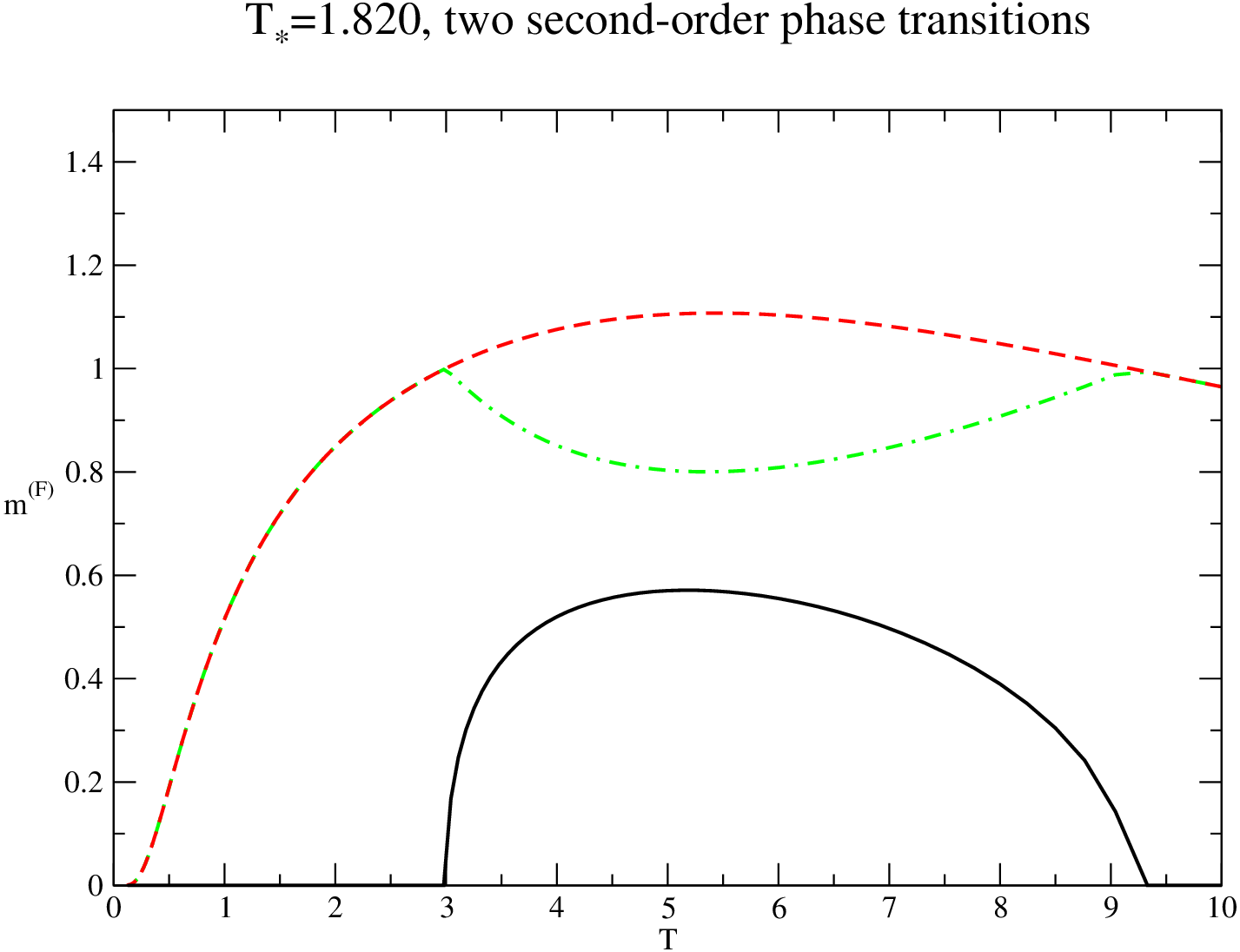}}
\caption{Magnetization (solid line), and curves of stability 
(dashed and dot-dashed lines) for the case
$c=1.4$, $J_0=-0.5$ and $J=10$. Here $T_{c1}=3.00$ and $T_{c2}=9.34$.} 
\label{sw_8}
\end{figure}
\begin{figure}
\epsfxsize=65mm \centerline{\epsffile{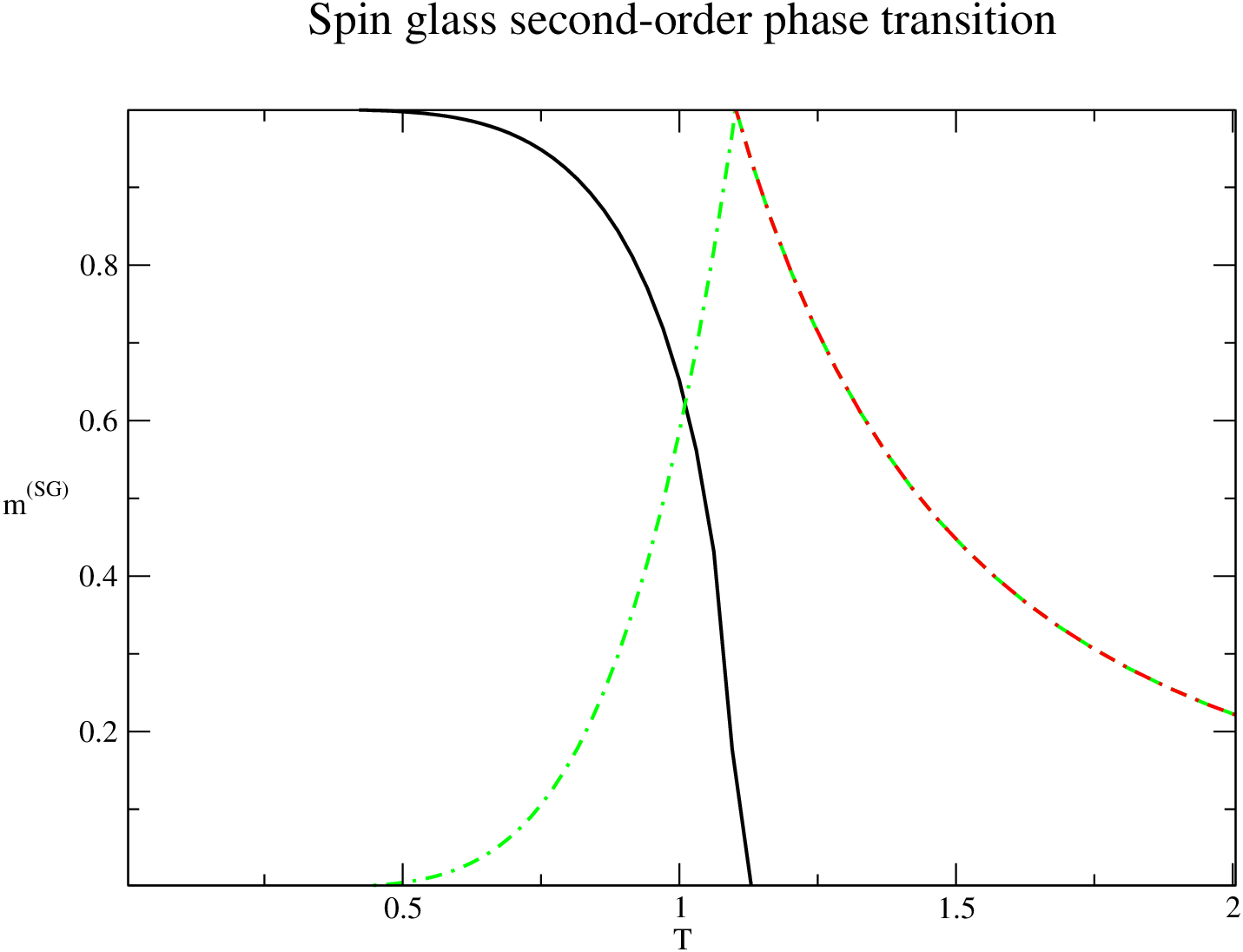}}
\caption{Spin glass order parameter (solid line), and curves of stability 
(dashed and dot-dashed lines) for the case
$c=0.5$, $J_0=1$ and $J=3/5/c$. Here $T_{c}=1.130$.} 
\label{sw_sg_1}
\end{figure}
\begin{figure}
\epsfxsize=65mm \centerline{\epsffile{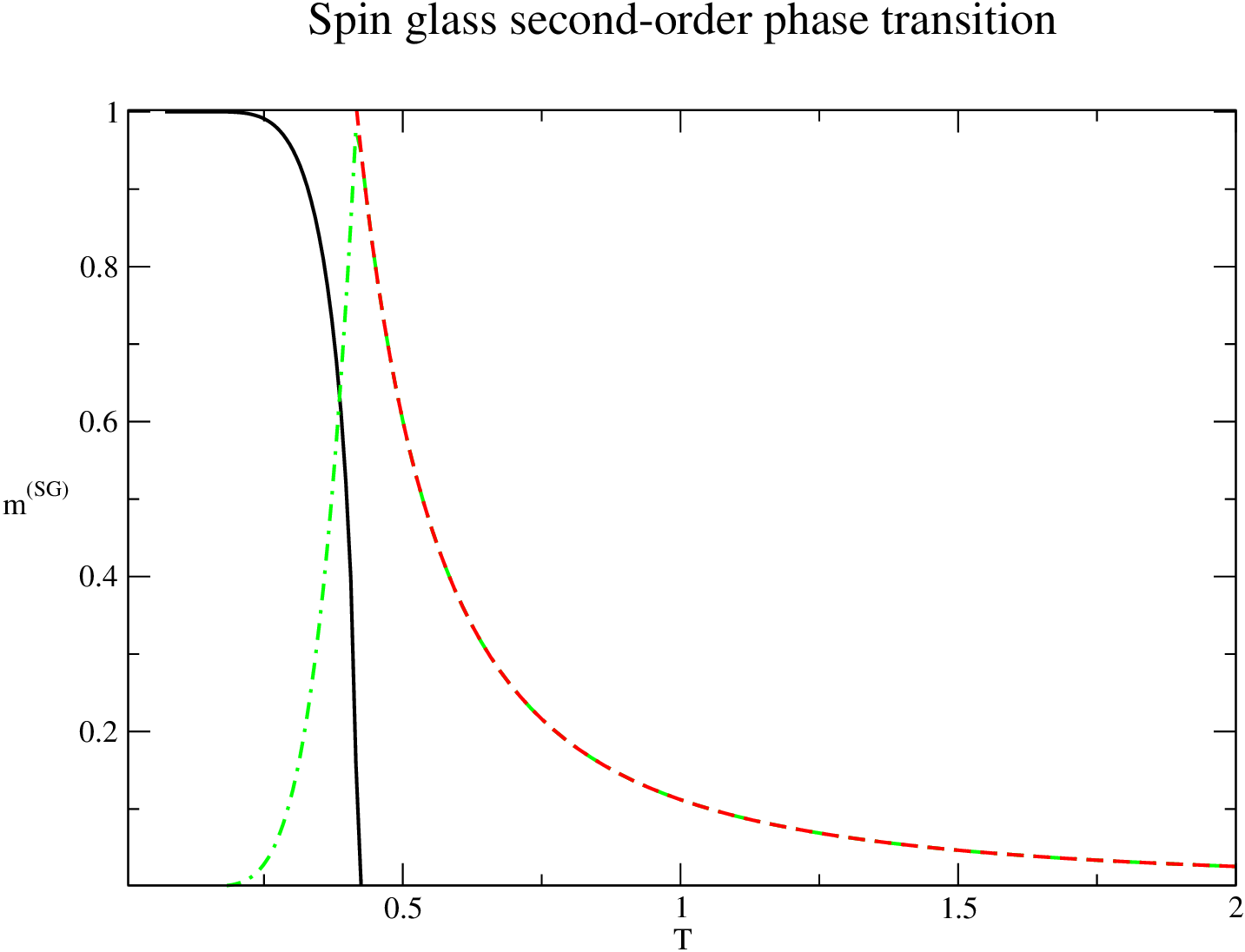}}
\caption{Spin glass order parameter (solid line), and curves of stability 
(dashed and dot-dashed lines) for the case
$c=10$, $J_0=0.25$ and $J=1/c$. Here $T_{c}=0.424$.} 
\label{sw_sg_2}
\end{figure}
\begin{figure}
\epsfxsize=65mm \centerline{\epsffile{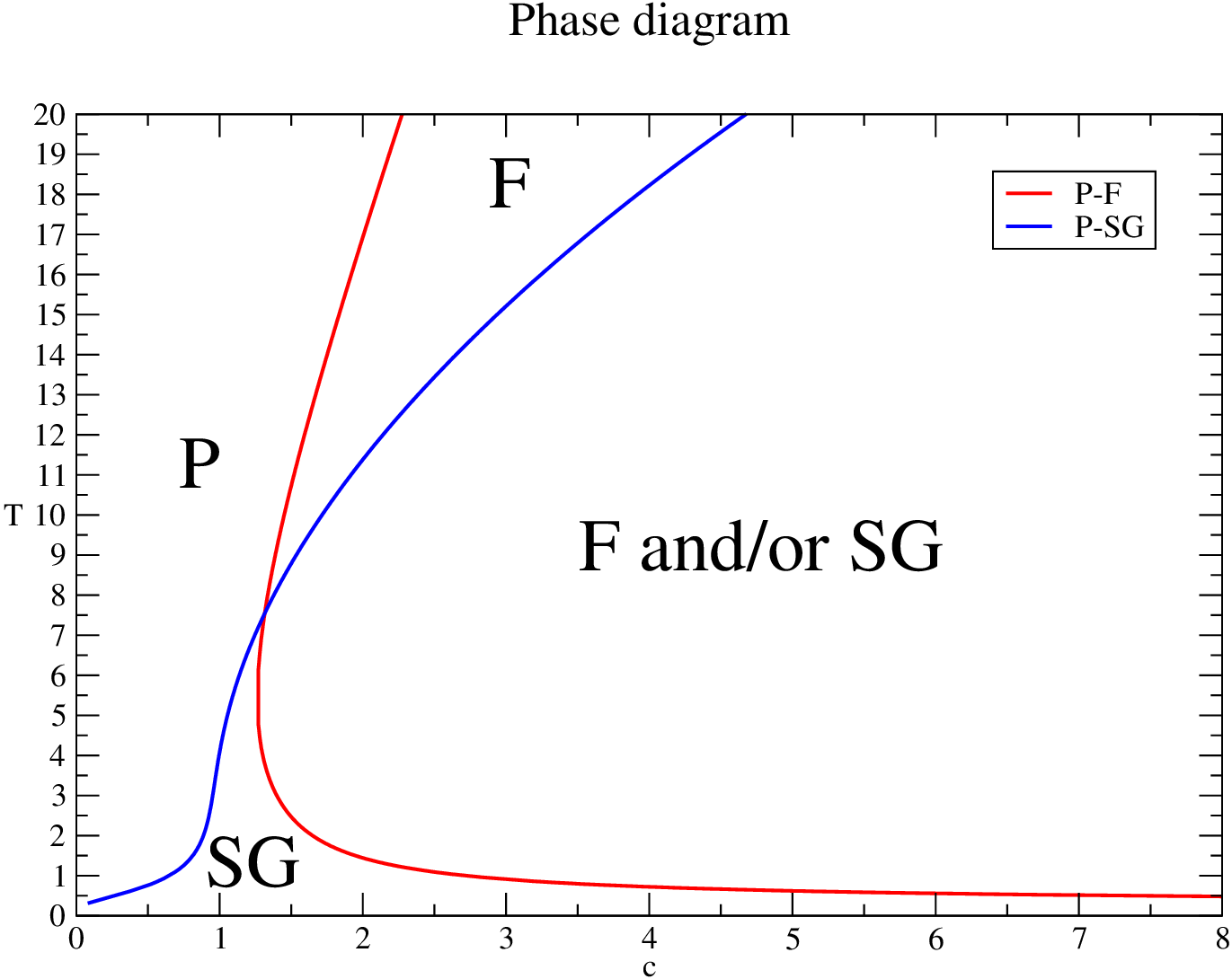}}
\caption{Phase diagram for the case considered in Fig. \ref{sw_7}
with the measure of Eq. (\ref{dPF}).}
\label{sw_pd1}
\end{figure}
\begin{figure}
\epsfxsize=65mm \centerline{\epsffile{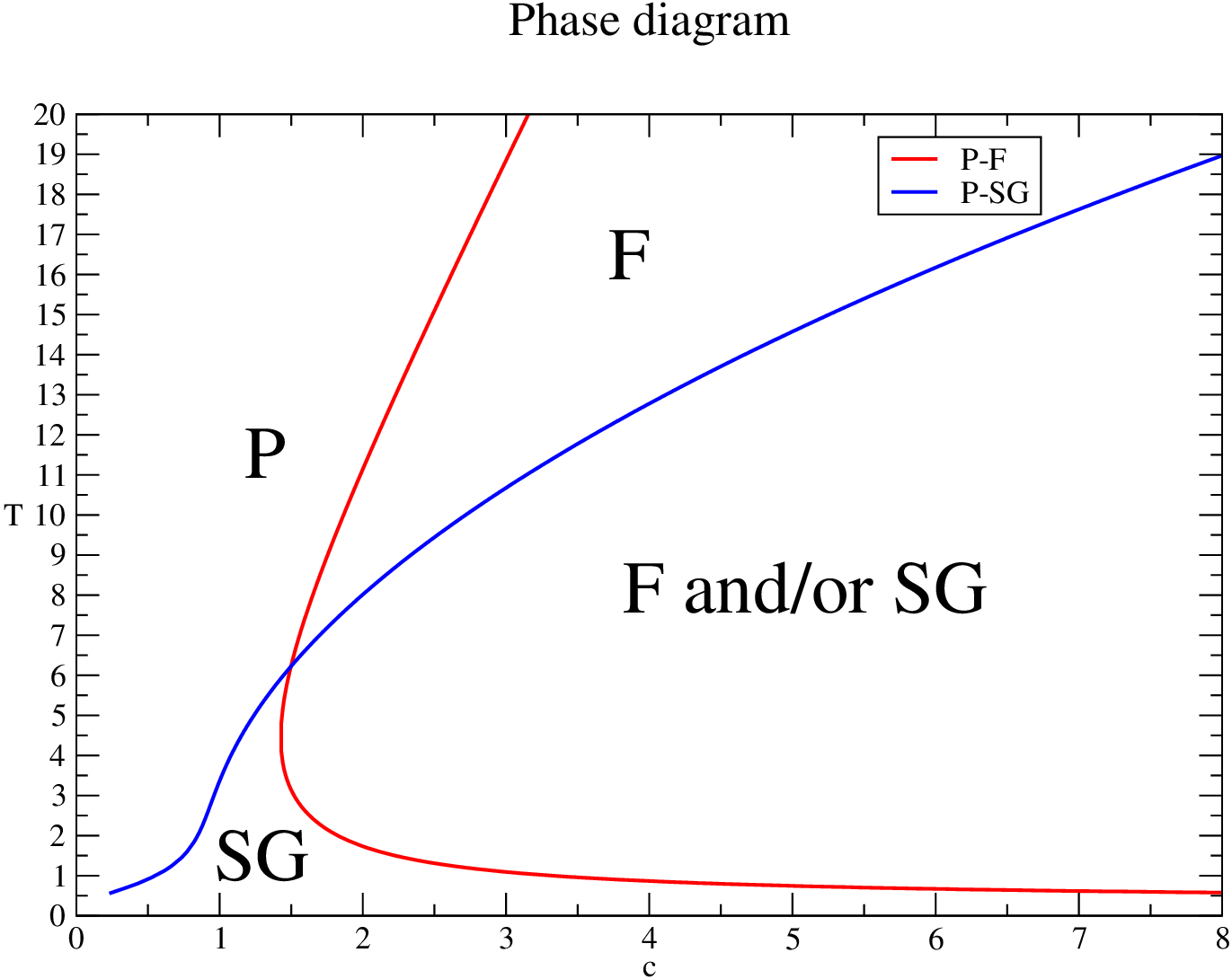}}
\caption{Phase diagram for the case considered in Fig. \ref{sw_8}
with the measure of Eq. (\ref{dPF}).}
\label{sw_pd2}
\end{figure}

In Figs. 1-10 we plot the 
magnetization $m^{(\mathrm{F})}$ (solid line), 
${\tilde{\chi}_0\left(\beta^{(\mathrm{F})}J_0^{(\mathrm{F})},0\right)}
\beta^{(\mathrm{F})}J^{(\mathrm{F})}$ (dashed line), and 
${\tilde{\chi}_0\left(\beta^{(\mathrm{F})}J_0^{(\mathrm{F})},
\beta J^{(\mathrm{F})}m^{(\mathrm{F})}\right)}
\beta^{(\mathrm{F})}J^{(\mathrm{F})}$ (dot-dashed line) for several cases
obtained by solving Eq. (\ref{1dl}) numerically with $\mathrm{\Sigma}$=F,
and by choosing
the stable solution minimizing $L^{(\mathrm{F})}(m)$ (see Eq. (\ref{THEOll})).
In all these examples we have chosen the measure (\ref{dPF}).
Figs. 1 and 2 concern two cases with $J_0>0$ so that one and only one
second-order phase transition is present. The input data of these two cases are
the same as those analyzed numerically in \cite{Niko} 
(note that in the model considered in \cite{Niko}, the long range
coupling $J$ is divided by $c$). As already stated in Sec. III, 
the self-consistent equations
become exact in the limit $c\to 0$, for second order phase transitions, 
and in the limit $c\to\infty$. 
Therefore, for the magnetization, by comparison with \cite{Niko}, 
in Fig. 1 and 2, where $c$ is relatively small and big, respectively, 
we see good agreement also below the critical temperature. 
 
Figs. 3-10 concern eight cases with $J_0<0$. 
As explained above, the critical behavior and the localization of
the critical temperatures is more complicated when $J_0<0$. In particular,
given $|J_0|$, if $c$ is not sufficiently high the solution $m^{(\mathrm{F})}=0$
remains stable at all temperatures and if it is also a leading solution,
no phase transition occurs.
Let us consider Eq. (\ref{1dm}). 
For $J_0<0$ the lhs of this equation has some maximum at a finite
value $\bar{\beta}$ given by
\begin{eqnarray}
\label{1dt}
\bar{\beta} J=\frac{1}{2}\log\left[\frac{1+\delta(r)}{1-\delta(r)}\right],
\end{eqnarray} 
where 
$r\equiv |J_0|/J$, and we have introduced
\begin{eqnarray}
\label{1du}
\delta(r)\equiv \sqrt{1+r^2}-r.
\end{eqnarray} 
Hence, we see that a sufficient condition 
for the solution $m^{(\mathrm{F})}=0$ to become unstable is that be
\begin{eqnarray}
\label{1dv}
c\left(\frac{1+\delta(r)}{1-\delta(r)}\right)^{r}\delta(r)\geq 1.
\end{eqnarray} 
Note that the above represents only a condition for the instability
of the solution $m^{(\mathrm{F})}=0$, but the true solution is the one
that is both stable and leading. In fact, when $J_0<0$,
a phase transition in general may be present also when Eq. (\ref{1dv}) 
is not satisfied and, correspondingly 
the possible critical temperatures will be not determined by Eq. (\ref{1dm}).
 
In Fig 3 we report a case with $J=1$, $J_0=-1$ ($r=1$) and a relatively low value of $c$,
$c=5$, so that no phase transition is present. Similarly, in Fig. 4
we report again a case in which no phase transition is present due to the fact
that here $r$ is relatively big, $r=1.1$.
It is interesting to observe that for $r=1$ Eq. (\ref{1dv}) requires a value
of $c$ greater than the limit value $c=3+2\sqrt{2}=5.8284...$.
In both of Figs. 5 and 6 
we report a case in which Eq. (\ref{1dv}) is still not satisfied,
but nevertheless two first-order phase transitions are present.    
In both of Figs. 7 and 8 we have one first- and one second-order phase transition. 
In both of Figs. 9 and 10 we have two second-order phase transitions. 
The critical temperature of second-order transition can be determined 
also by Eq. (\ref{1dm}). 
In the top of Figs. 5-10 we write the discriminant 
temperature $T_*=4|J_0|/\log(3)$ below which
a phase transition (if any) is first-order (see Eqs. (\ref{1dg})-(\ref{1di})) 
(see Sec. IIIB of part I).

Finally in Figs. 11 and 12 we plot the 
spin glass order parameter $m^{(\mathrm{SG})}$ (solid line), 
${\tilde{\chi}_0\left(\beta^{(\mathrm{SG})}J_0^{(\mathrm{SG})},0\right)}
\beta^{(\mathrm{SG})}J^{(\mathrm{SG})}$ (dashed line), and 
${\tilde{\chi}_0\left(\beta^{(\mathrm{SG})}J_0^{(\mathrm{SG})},
\beta J^{(\mathrm{SG})}m^{(\mathrm{SG})}\right)}
\beta^{(\mathrm{SG})}J^{(\mathrm{SG})}$ (dot-dashed line) 
obtained by solving Eq. (\ref{1dl}) numerically with $\mathrm{\Sigma}$=SG.
In these two examples we have chosen the measure (\ref{dPSG}) and,
for $c$, $J$, and $J_0$, we have considered the same parameters 
of Figs. 1 and 2 of the ferromagnetic case.

Note that, unlike the P-F critical surface, the P-SG critical
surface does not depend on the parameter $p$ entering in Eq. (\ref{dPSG}).
For the reciprocal stability between the P-F and the P-SG critical 
surface we remind the reader to the general rules of Sec. IIID of part I 
(see cases \textbf{(1)} and \textbf{(3)}) which, for $J_0\geq 0$, 
reduce to the results reported in Sec. 6.1 of the Ref. \cite{Niko}. 
Here we stress just that, if $J_0\geq 0$, for $p\leq 0.5$, only the P-SG transition is possible.
However, when $J_0<0$ and $c$ is not sufficiently large, 
the SG phase may be the only stable phase even when $p=1$. 
In fact, although when $J_0<0$ the solution $m^{(\mathrm{F})}$
may have two P-F critical temperatures, in general, if the P-SG
temperature is between these, we cannot exclude that the solution 
$m^{(\mathrm{SG})}$ starts to be the leading solution at sufficiently
low temperatures. In Figs. \ref{sw_pd1} and \ref{sw_pd2},
on the plane $(T,c)$, we plot the phase diagrams
corresponding to the cases of Figs. \ref{sw_7} and \ref{sw_8}, respectively.
Theses phase diagrams are obtained by solving Eq. (\ref{THEOg}) supposing that
here, as in the cases of Figs. \ref{sw_7} and \ref{sw_8}, where
$c=1.4$ and $c=0.5$, respectively, the P-F transition is always second-order.
We plan to investigate in more detail the phase diagram in future works.

\section{Small-world spherical model in arbitrary dimension $d_0$}
In this section we will analyze the case in which the unperturbed model is
the spherical model built up over
a $d_0$-dimensional lattice $\mathcal{L}_0$ (see \cite{Baxter} and references
therein) \cite{Note}.
In this case the $\sigma$'s are continuous ``spin'' variables ranging
in the interval $(-\infty,\infty)$ subjected to the sole constrain 
$\sum_{i\in\mathcal{L}_0}\sigma_i^2=N$, however our theorems and formalism can
be applied as well and give results that, within the same limitations
prescribed in Sec. III, are exact. 

Following \cite{Baxter}, for the unperturbed model we have
\begin{eqnarray}
\label{1s}
-\beta f_0 (\beta J_0,\beta h)=
\frac{1}{2}\log\left(\frac{\pi}{\beta J_0}\right)+
\phi\left(\beta J_0,\beta h,\bar{z}\right),
\end{eqnarray} 
\begin{eqnarray}
\label{2s}
m_0 (\beta J_0,\beta h)=\frac{\beta h}{2\beta J_0 \bar{z}},
\end{eqnarray} 
where
\begin{eqnarray}
\label{3s}
\phi\left(\beta J_0,\beta h,z\right)=\beta J_0 d_0 + \beta J_0 z - \frac{1}{2}g(z)+
\frac{\left(\beta h\right)^2}{4\beta J_0 z},
\end{eqnarray} 
\begin{eqnarray}
\label{4s}
g(z)&=&\frac{1}{(2\pi)^{d_0}}\int_0^{2\pi}\ldots \int_0^{2\pi}
d\omega_1\ldots d\omega_{d_0}
\nonumber \\
&\times& \log\left[d_0+z-\cos(\omega_1)-\ldots-\cos(\omega_{d_0})\right],
\end{eqnarray} 
and $\bar{z}=\bar{z}(\beta J_0,\beta h)$ is the (unique) solution of 
the equation $\partial_{z}\phi\left(\beta J_0,\beta h,z\right)=0$:
\begin{eqnarray}
\label{5s}
\beta J_0-\frac{\left(\beta h\right)^2}{4\beta J_0 \bar{z}^2}=\frac{1}{2}g'(\bar{z}),
\end{eqnarray} 
from which it follows the equation for $m_0$ 
\begin{eqnarray}
\label{6s}
\beta J_0\left(1-m_0^2\right)=\frac{1}{2}g'\left(\frac{\beta h}{2\beta J_0 m_0}\right).
\end{eqnarray} 
The derivative $g'$ can in turn be expressed as
\begin{eqnarray}
\label{7s}
g'(z)=\int_0^{\infty}e^{-t(z+d_0)}\left[\mathcal{J}_0\left(it\right)\right]^{d_0} dt,
\end{eqnarray} 
$\mathcal{J}_0\left(it\right)$ being the usual Bessel function whose
behavior for large $t$ is given by
\begin{eqnarray}
\label{8s}
\mathcal{J}_0\left(it\right) =
\frac{e^t}{(2\pi t)^{\frac{1}{2}}}\left(1+\mathop{O}\left(\frac{1}{t}\right)\right).
\end{eqnarray} 

The critical behavior of the unperturbed system depends on the values of
$g'(z)$ and $g''(z)$ near $z=0$. 
It turns out that for $d_0\leq 2$ one has $g'(0)=\infty $ and there is no
spontaneous magnetization, whereas for $d_0>2$ one has $g'(0)<\infty $ and at
$h=0$ the unperturbed system undergoes a second-order phase transition with
magnetization given by Eq. (\ref{6s}) which, for $\beta$ above
$\beta_{c0}$, becomes
\begin{eqnarray}
\label{9s}
m_0(\beta J_0,0)=\sqrt{1-\frac{\beta_{c0}}{\beta}},
\end{eqnarray} 
where the inverse critical temperature $\beta_{c0}$ is given by
\begin{eqnarray}
\label{10s}
\beta_{c0}J_0=\frac{1}{2}g'\left(0\right).
\end{eqnarray} 
Furthermore, it turns out that for $d_0\leq 4$ one has $g''(0)=\infty $, 
whereas for $d_0>4$ one has $g''(0)<\infty $. This reflects on the
critical exponents $\alpha$, $\gamma$ and $\delta$, which take the classical
mean-field values only for $d_0>4$.

According to sec. III, to solve the random model - for simplicity - 
at zero external field, we have to perform the
effective substitutions $\beta J_0\to \beta J_0^{(\Sigma)}$ and
$\beta h\to \beta J^{(\Sigma)} m^{(\Sigma)}$ in the above equations.
From Eqs. (\ref{2s}), (\ref{5s}) and (\ref{6s}), we get immediately: 
\begin{eqnarray}
\label{11s}
\bar{z}^{(\Sigma)}=
\frac{\beta J^{(\Sigma)}}
{2\beta J_0^{(\Sigma)}},
\end{eqnarray} 
the equations for
inverse critical temperature $\beta_c^{(\Sigma)}$  
\begin{eqnarray}
\label{12s}
\beta_c^{(\Sigma)} J_0^{(\Sigma)}=
\frac{1}{2}g'\left(
\frac{\beta_c^{(\Sigma)}J^{(\Sigma)}}
{2\beta_c^{(\Sigma)} J_0^{(\Sigma)}}\right),
\end{eqnarray} 
and the magnetizations 
$m^{(\Sigma)}$
\begin{eqnarray}
\label{13s}
m^{(\Sigma)}=\left\{
\begin{array}{l}
\sqrt{1-
\frac{1}{2\beta J_0^{(\Sigma)}}
g'\left(\frac{\beta J^{(\Sigma)}}
{2\beta J_0^{(\Sigma)}}\right)},\quad \beta>\beta_c^{(\Sigma)},
\\
0,\quad \beta<\beta_c^{(\Sigma)}\geq 0.
\end{array}
\right.
\end{eqnarray} 
Note that, as it must be from the general result of Sec. IIIB of part I,
unlike the unperturbed model, as soon as the connectivity 
$c$ is not zero, Eq. (\ref{12s}) has always a finite
solution $\beta_c^{(\Sigma)}$, independently on the dimension $d_0$.
In fact, one has a finite temperature second-order 
phase transition even for $d_0\to 0^+$ where
from Eq. (\ref{7s}) we have  
\begin{eqnarray}
\label{7sb}
g'(z)=\frac{1}{z}, \qquad (d_0=0)
\end{eqnarray} 
so that the equations for the critical temperature (\ref{12s}) become
\begin{eqnarray}
\label{12sb}
\beta_c^{(\Sigma)} J^{(\Sigma)}=1, \qquad (d_0=0)
\end{eqnarray} 
which, as expected, coincide with Eqs. (\ref{VBh}) and (\ref{VBi}) 
of the Viana-Bray model.

Similarly, unlike the unperturbed model, in the random model
all the critical exponents take the classical mean-field values,
independently on the dimension $d_0$. In the specific case of
the spherical model, this behavior is due to the fact
that $g'(z)$ and $g''(z)$ can be singular only at $z=0$
but, as soon as the connectivity $c$ is not zero,
there is an effective external field $\beta J^{(\Sigma)} m^{(\Sigma)}$ 
so that $\bar{z}^{(\Sigma)}$ is not zero.
For the critical behavior, 
the dependence on the dimension $d_0$ reflects only in 
the coefficients, not on the critical exponents.
In particular, concerning the argument of the square root
of the rhs of Eq. (\ref{13s}), 
by expanding in the reduced temperature $t^{(\Sigma)}$, 
for $|t^{(\Sigma)}|\ll 1$ we have
\begin{eqnarray}
\label{14s}
&& 1-\frac{1}{2\beta_c^{(\Sigma)} J_0^{(\Sigma)}}
g'\left(\frac{\beta_c^{(\Sigma)}J^{(\Sigma)}}
{2\beta_c^{(\Sigma)} J_0^{(\Sigma)}}\right)=\nonumber \\ 
&& B^{(\Sigma)} t^{(\Sigma)}+\mathop{O}(t^{(\Sigma)})^2,
\end{eqnarray} 
where
\begin{widetext}
\begin{eqnarray}
\label{15s}
B^{\mathrm{(F)}}&=&-1+
\frac{1}{2\beta_c^{\mathrm{(F)}} J_0^{\mathrm{(F)}}}
g''\left(\frac{\beta_c^{\mathrm{(F)}}J^{\mathrm{(F)}}}
{(2\beta_c^{\mathrm{(F)}} J_0^{\mathrm{(F)}})^2}\right)
\nonumber \\ \times
&& \left(c\int d\mu(J_{i,j}) \left(1-
\tanh^2(\beta_c^{\mathrm{(F)}} J_{i,j})\right)\beta_c^{\mathrm{(F)}} J_{i,j}
- c\int d\mu(J_{i,j}) \tanh(\beta_c^{\mathrm{(F)}} J_{i,j})\right),
\end{eqnarray} 
\begin{eqnarray}
\label{16s}
B^{\mathrm{(SG)}}&=&\frac{1}{2\beta_c^{\mathrm{(SG)}}J_0^{\mathrm{(SG)}}}
\left[-4\frac{\tanh(\beta_c^{\mathrm{(SG)}}J_0)\beta_c^{\mathrm{(SG)}}J_0}
{1+\tanh^2(\beta_c^{\mathrm{(SG)}}J_0)}+
g''\left(\frac{\beta_c^{\mathrm{(SG)}}J^{\mathrm{(SG)}}}
{(2\beta_c^{\mathrm{(SG)}} J_0^{\mathrm{(SG)}})^2}\right) \right.
\nonumber \\ \times
&& \left(2c\int d\mu(J_{i,j}) \left(1-
\tanh^2(\beta_c^{\mathrm{(SG)}} J_{i,j})\right)
\tanh(\beta_c^{\mathrm{(SG)}} J_{i,j})\beta_c^{\mathrm{(SG)}} J_{i,j} \right.
\nonumber \\ 
&& \left. \left. - 4c\int d\mu(J_{i,j}) \tanh^2(\beta_c^{\mathrm{(SG)}} J_{i,j})
\frac{\tanh(\beta_c^{\mathrm{(SG)}}J_0)\beta_c^{\mathrm{(SG)}}J_0}
{\left(1+\tanh^2(\beta_c^{\mathrm{(SG)}}J_0)\right)
\beta_c^{\mathrm{(SG)}}J^{\mathrm{(SG)}}}
\right)\right],
\end{eqnarray} 
\end{widetext}
so that from Eqs. (\ref{13s}) and (\ref{14s}) 
for the critical behavior of the magnetizations we get explicitly the
mean-field behavior:
\begin{eqnarray}
\label{17s}
m^{(\Sigma)}=\left\{
\begin{array}{l}
\sqrt{B^{(\Sigma)} t^{(\Sigma)}~}+\mathop{O}(t^{(\Sigma)}),\quad t^{(\Sigma)}<0,
\\
0,\quad t^{(\Sigma)}\geq 0.
\end{array}
\right.
\end{eqnarray} 

\section{Conclusions}
We have applied a novel method to solve analytically
several small-world models of interest defined over
an underlying lattice $\mathcal{L}_0$ of dimension $d_0=0,1$,and $\infty$,
corresponding to an ensemble of non interacting units
(spins, dimers, etc...), the one dimensional chain, and the spherical
model, respectively. The long-range couplings may also have
an additional own disorder leading, in particular, to spin glass phases.
The simplicity of the method allows us 
to find very easily (just a few line equations) the critical
surfaces and the order parameters. In our method the former are exact
whereas the latter provide an approximation to the exact order
parameters which, in the absence of frustration, 
become exact in the limit $c\to 0$, when the transition is of second order,
and in the limit $c\to\infty$.
In particular, we have studied in detail the small-world defined 
over the one dimensional chain with positive and 
negative short-range couplings, showing explicitly how, in the second case, multicritical
points with first- and second-order phase transition arise.
Finally, the small-world spherical model - an exact solvable model
(in our approach) with continuous spin variables - has provided us 
an interesting testing bench case to study what happens as $d_0$
changes continuously from 0 to $\infty$. As expected from general grounds, 
unlike the non random version of the model,
the small-world model presents always a finite temperature phase transition,
even in the limit $d_0\to 0^+$.
This latter result - besides to be consistent with what we have found 
in the $d_0=0$-dimensional discrete models -
is easily and physically explained by the method.
In fact, it consists on mapping the small-world model (a random model)
to a corresponding non random model (no long-range bonds) but immersed
in an effective uniform external field 
%
which is active as soon as the 
added random connectivity $c$ is not zero 
(see Eqs. (\ref{THEOa})-(\ref{THEOc}) ).

In this paper we have considered small-world models which - by the method -
turn out to be the simplest ones (cases for which the unperturbed model
is analytically feasible). 
Many interesting variants of these models can
be as well considered and are still analytically solvable by the method.
However, the method can be also applied numerically to study other more complex 
small-world models for which the corresponding unperturbed model is 
not analytically solvable. 
In fact, the numerical complexity in solving such a small-world model,  
will be still as feasible as a non random model immersed in a uniform external field.


\begin{acknowledgments}
This work was supported by the FCT (Portugal) grants
SFRH/BPD/24214/2005, pocTI/FAT/46241/2002 and
pocTI/FAT/46176/2003, and the Dysonet Project.
We thank A. L. Ferreira for useful discussions.
\end{acknowledgments}




\begin{thebibliography}{10}
\bibitem{Note} Note that from the point of view of the statistical mechanics the
spherical model is an infinite dimensional model, however we will keep on to
reserve the symbol $d_0$ for the dimension of $\mathcal{L}_0$.
\end{thebibliography}
\end{document}